\documentclass[prb,aps,twocolumn,10pt,superscriptaddress,notitlepage,longbibliography]{revtex4-2}

\usepackage{graphicx}
\usepackage{amsmath,mathtools}
\usepackage{amssymb}
\usepackage{epstopdf}
\usepackage{siunitx}
\usepackage{color}
\usepackage{bm}
\usepackage[ngerman,english]{babel}

\usepackage{hyperref}
 \hypersetup{
     colorlinks=true,
     linkcolor=blue,
     filecolor=blue,
     citecolor = magenta,      
     urlcolor=red,
     }

\usepackage{braket}

\usepackage{letltxmacro}
\LetLtxMacro{\ORIGselectlanguage}{\selectlanguage}
\makeatletter
\DeclareRobustCommand{\selectlanguage}[1]{%
  \@ifundefined{alias@\string#1}
    {\ORIGselectlanguage{#1}}
    {\begingroup\edef\x{\endgroup
       \noexpand\ORIGselectlanguage{\@nameuse{alias@#1}}}\x}%
}
\newcommand{\definelanguagealias}[2]{%
  \@namedef{alias@#1}{#2}%
}

\makeatother

\definelanguagealias{en}{english}
\definelanguagealias{EN}{english}
\definelanguagealias{eng}{english}
\definelanguagealias{de}{ngerman}

\graphicspath{{./figs/}}

\begin{document}

\title{Strong coupling regime and hybrid quasinormal modes of a 
single plasmonic resonator coupled to a TMDC monolayer}

\date{\today}
\author{Chelsea Carlson}
\address{Department of Physics, Queen's University, Kingston, Ontario, Canada, K7L 3N6}
\author{Robert Salzwedel}
\address{Institute of Theoretical Physics, Nonlinear Optics and Quantum Electronics, Technische Universität Berlin, 10623 Berlin, Germany}
\author{Malte Selig}
\address{Institute of Theoretical Physics, Nonlinear Optics and Quantum Electronics, Technische Universität Berlin, 10623 Berlin, Germany}
\author{Andreas Knorr}
\address{Institute of Theoretical Physics, Nonlinear Optics and Quantum Electronics, Technische Universität Berlin, 10623 Berlin, Germany}
\author{Stephen Hughes}
\address{Department of Physics, Queen's University, Kingston, Ontario, Canada, K7L 3N6}

\begin{abstract}
     We  present a rigorous photonic mode model to describe
    the  strong coupling between a monolayer of $\rm MoSe_2$ and a  single gold nanoparticle. The onset of strong coupling is quantified by computing the three-dimensional hybrid quasinormal modes of the combined structure, allowing one to accurately model light-matter interactions without invoking the usual phenomenological theories of strong coupling.
    We explore the hybrid quasinormal modes as a function of gap size and temperature and find spectral splittings in the range of around 80-110 meV, with no fitting parameters for the material models. We also show how the hybrid modes exhibit Fano-like resonances and quantify the complex poles of the hybrid modes as well as the Purcell factor resonances from embedded dipole emitters. These effects cannot be described with the usual heuristic normal mode theories.
\end{abstract}

\maketitle

\section{Introduction}
\label{sec:0}

One common goal of nano-optics is to create photonic cavity modes that can significantly enhance light-matter interactions, leading to new capabilities in sensing and quantum optics. While much success has been achieved with semiconductor cavity structures~\cite{Vahala2003}, recently substantial attention has been devoted to metallic nanoparticles (MNPs), inspired  by their ability to trap light in sub-wavelength spatial scales and allow a broadband enhanced coupling. This can help enable new regimes in quantum plasmonics~\cite{TameQuantumPlasmonics,QuantumPlasmonicsReview}, including the experimental demonstrating of strongly coupled single molecules and MNPs at room temperature~\cite{Chikkaraddy2016}. 

Recently, there has been much interest to increase the interaction strengths even further by coupling to 2D  
semiconductors, such as transition metal dichalcogenides (TMDCs).
Monolayers of TMDCs are direct gap semiconductors with strong light matter interactions which makes them promising for optoelectronic applications~\cite{xiao2012coupled,kormanyos2015k}.  The reduced dimensionality of these atomically thin materials leads to a boost of the Coulomb interaction which is responsible for the formation of tightly bound electron hole pairs (excitons)~\cite{chernikov2014exciton}. The large excitonic binding energies make TMDCs a promising 
platform to study exciton physics~\cite{steinhoff2017exciton,katsch2020exciton}. 
The TMDC excitons form separate, spectrally well-isolated resonances below the band gap and can couple radiatively to MNPs if their dielectric surrounding is properly designed. This allows one to also achieve signatures
of strong coupling from near resonant TMDC-MNP hybrid systems which gives rise to large Rabi splittings.

To date, a variety of MNPs have been used to observe  signatures of strong coupling 
between MNPs and TMDCs, including single MNP nanorods~\cite{kern-nanoantenna-2015,zheng-manipulating-2017,wen-room-temperature-2017}, bi-pyramids~\cite{struhrenberg-strong-2018}, disks~\cite{geisler-single-2019}, spheroids~\cite{kleeman-strong-2017}, dimers~\cite{liu-strong-2020}, and arrays of MNPs~\cite{abid-temperature-2017,abid-surface-2018,bisht-collective-2019}. Measurements are often performed using reflection/transmission from a low-powered laser or dark-field scattering~\cite{kern-nanoantenna-2015}, and are typically performed at room temperature. Impressive spectral Rabi splittings of 500~meV have been achieved using a microcavity coupled to a TMDC sheet with arrays of gold disks~\cite{bisht-collective-2019} (although the splitting arises primarily from the classical mode splitting between the cavity and MNP array), while the average Rabi splittings in recent works range from approximately 50-150~meV. 
Most if not all of these works are in the regime
of so-called ``normal mode splitting,'' and thus the physics of the splitting can
be well described classically \cite{schneider-two-dimensional-2018}. 
Despite numerous experimental observations, there are no rigorous 3D models
in the literature to properly assess the strong coupling regime in these material systems, which is significantly more challenging than typical planar structures, e.g., with quantum wells~\cite{PhysRevLett.69.3314}, since the MNPs break translational invariance. The terminology of ``normal mode splitting'' is also not appropriate, as the MNPs on their own do not support such modes, requiring a more rigorous mode theory from the beginning.

Besides normal mode splitting with planar-like cavity systems,
 strong coupling phenomena of single quantum dots (which naturally break translational invariance) have been reported
in a number of semiconductor microcavity systems~\cite{reithmaier-strong-2004,Yoshie2004,peter-exciton-2005},
where the emitter-cavity coupling rate is defined through
 $g = (\pi e^2 f)/(4\pi\epsilon_r\epsilon_0m_0V_m)^{1/2}$, with $f$ the oscillator strength, $m_0$  the free electron mass, and $V_c$  the cavity mode volume. In these systems, experimental Rabi splittings up to 140~$\mu$eV have been reported, and  the description of $g$ as a cavity-emitter coupling rate is more appropriate since it is describing a single quantum emitter and the light field within a dipole approximation; the single mode approximation is also excellent here since 
 the cavity system is usually very low loss, yielding substantial
quality factors ($Q>1000$). The interpretation of these emitter-cavity systems in the strong coupling regime is clear since the emitter-cavity coupling rate is larger than any dissipation rates. However, since the upper limit to $g$ in these systems is bound by the diffraction limit, most, if not all, of the single quantum dot experiments in this field have to work at very low temperatures. 
    
    Despite the similarities between
    point dipole emitters in cavities
    and TMDC-plasmon systems, to our knowledge, no one has reported a first-principles
electromagnetic
calculation of the underlying hybrid modes for strongly dissipating metal structures using material models with fully microscopically determined parameters for the TMDC. 
The understanding of strong coupling is also essential to develop
mode quantization theories, which can then allow true signatures of quantum optical effects in these material systems.

In this work, we introduce a rigorous electromagnetic theory and model to describe the regime of
strong coupling in TMDC-MNP systems, including a full analysis of the underlying hybrid modes. In particular, we develop
a quasinormal mode (QNM) theory of such hybrid systems, enabling a  full 3D description of the modes, light-matter interactions,  and the strong coupling behavior of these modes. Quasinormal modes are known to be highly accurate for describing such hybrid material systems, and have recently shown much success
in modelling a wide range of resonators in nano-optics and plasmonics~\cite{kristensen-modes-2014,PhysRevA.98.043806,PhysRevB.94.235438,lalanne-light-2018,kristensen-modeling-2020}. 
Here  we explore the QNMs of a MoSe$_2$ monolayer TMDC encapsulated in hBN coupled to a gold nanorod MNP. 
By tuning the dimensions of the MNP to get the desired resonance frequency, we show strong coupling between the second lowest mode
of the MNP, which is the next mode in frequency space from the fundamental mode, and the TMDC for various gap sizes between the two structures. 

We then compare the QNM analysis to the full-dipole numerical calculations (with no approximations), and show excellent agreement to justify the validity of the QNM analysis
over a wide range of frequencies and 
 spatial points near the resonators. We also compare the optical response for selected temperatures of 4, 77, and 300~K with microscopically calculated material parameters~\cite{selig2016excitonic,khatibi2018impact} and show the anti-crossing behavior when detuning the resonances by varying the size of the MNP, which is a clear signature of strong-coupling.  We then compare the splitting observed from the spectral Purcell response and the splitting between the complex poles of the hybrid QNMs, with no fitting parameters.  Lastly,
we also study how the transmission spectra from an embedded dipole differs from the near field Purcell factor, which show complimentary but different signatures of the strong coupling regime. The dipole transmission requires the full two-space-point Green function and accounts for spatial quenching between the dipole and the detection point, while the Purcell factor is a measure of the projected local density of states at the dipole location, which is also contained within the Green function with equal space points. Conveniently, we construct an analytical solution to the full two-space-point Green function through an analytical expansion of the hybrid QNMs, which  critically contains the QNM phase,
and can be applied to a wide range of problems in nanophotonics.


\section{Summary of common models  used in the literature}

We first summarize some common models and   results reported in 
the literature, and discuss the various formalisms currently being used
to explain the TMDC-MNP strong coupling behavior.
For this specific work we will refer to ``strong coupling''
in the semiclassical sense of QNM spectral splitting when the two hybrid modes are suitably coupled together.
As is well known, strong coupling
in the linear regime can also be explained from
the classical perspective of linear dispersion theory~\cite{PhysRevLett.64.2499}, and this is essentially the regime we also explore below, though we adopt a rigorous QNM approach.

We begin by introducing
 the usual textbook model of
coupled harmonic oscillators, also commonly used to
  describe  strong coupling and ({\it sic}) ``normal mode splitting'' in TMDC-MNP systems~\cite{schneider-two-dimensional-2018,hu-recent-2020}:
\begin{equation}
    \begin{bmatrix} \dot{\mathbf{P}} \\ \dot{\mathbf{E}} \end{bmatrix} = 
     \begin{bmatrix} (\omega_{x} - i\gamma_{x}) & g \\ g & (\omega_{c} - i\gamma_{c}) \end{bmatrix}
    \begin{bmatrix} {\mathbf{P}} \\ {\mathbf{E}} \end{bmatrix},
\end{equation}
where $\mathbf{P}$ is the polarizability of the TMDC exciton due to the dominant exciton of interest, $\mathbf{E}$ is the electric field of the interacting cavity mode, $g$ is the exciton-cavity coupling rate, $\gamma_{x, c}$ is the phenomenological decay rate of the TMDC exciton or cavity mode, and $\omega_{ex, c}$ is the resonant frequency of the TMDC or cavity. The fields $\mathbf{P}$ and $\mathbf{E}$ are both 3D vectors in general, but the excitons live primarily in-plane to the TMDC, so these vectors are effectively 2D, although out-of-plane excitons with comparably weak radiative interaction can exist using stacked monolayer designs \cite{kleeman-strong-2017}.  

The strong coupling regime here
 characterizes the coupled mode splitting of some spectral signature such as the emission spectra, which can be seen as representing classical oscillations of particles and fields in the time domain (if adopting a two level atom model for one of the oscillators).
This also allows one to construct a quantized field theory in order to invoke the quantum language of strong coupling
(i.e., vacuum Rabi oscillations) to connect to 
quantum optics formalisms like the Jaynes-Cummings (JC) model.
However, in contrast to spatially extended systems such as MNP-TMDC hybrids, the JC 
model (or quantum Rabi model, if a rotating wave approximation is not made~\cite{frisk-kockum-ultrastrong-2019}) either assumes point coupling (such as in atom cavity-QED), or constant coupling as a funbction of space, such as with quantum wells in DBR (distributed Bragg reflector) mirrors.

Phenomenologically, as a first approximation, the eigenfrequencies of this simplified hybrid system can be found by assuming a harmonic time-dependence for $\mathbf{P}$, $\mathbf{E}$ $\propto \exp{(\pm i\omega t})$. When the system is on resonance, such that $\omega_0 \equiv \omega_c = \omega_{x}$, and
assuming that $\gamma_{x},~\gamma_{c}\ll g,~\omega_c,~\omega_{x}$, the complex eigenfrequencies are readily obtained as:
\begin{equation}
    \omega_\pm = \omega_0 - \frac{i(\gamma_c+\gamma_{ex}) }{2} \pm \frac{\Omega_R}{2},
\end{equation}
where $\hbar\Omega_R\equiv \hbar\sqrt{4g^2 - (\gamma_{ex}-\gamma_c)^2}$ is the spectral mode splitting. Thus, the coupling constant can be determined by the observed splitting, using
\begin{equation}
    g = \frac{\sqrt{\Omega_R^2 + (\gamma_c-\gamma_{ex})^2}}{2},
    \label{eq:g}
\end{equation}
which can be thought of as a JC type cavity-emitter coupling rate, though this is usually defined for a single two-level atom or point dipole emitter~\cite{pelton-strong-2019}. 

The usual criterion for observing the strong coupling regime, for vacuum dynamics or linear excitation, is often given by the condition:
$ \Omega_R = 2 g > {\vert \gamma_c-\gamma_{ex}\vert}$.
However, for large damping rates
of the cavity mode,
a more accurate measure is given by
\cite{schneider-two-dimensional-2018,zheng-manipulating-2017,wen-room-temperature-2017}
\begin{equation}
    \Omega_R> \gamma_{x}+\gamma_{c},
    \label{eq:g_b}
\end{equation}
 but even this criterion is only a crude approximation
 for spatially extended TMDC-MNP systems.
Clearly one needs to understand the underlying properties of the modes before defining what is meant here by strong coupling, especially for the development of a quantum model.
In particular, at best, $g$ is an {\it effective coupling constant}, 
but it does not tell us anything about the spatially dependent coupling. 

For the strong coupling regime, in addition 
to satisfying the approximate
relation, $\Omega_R > \gamma_x + \gamma_c$,
the other usual assumption is that
 $g/\omega_0<0.1$, otherwise the rotating-wave approximations used in the derivation of the JC models does not work. Above this threshold, we enter the so-called ``ultrastrong'' coupling (USC)
 regime~\cite{frisk-kockum-ultrastrong-2019,forn-ultrastrong-2019}. 
However, at the semi-classical level, such effects are in fact easy to account for, namely we do not have to invoke a rotating wave approximation in the theory.

To date,
almost all works to describe strong coupling between TMDCs and MNPs  in the literature use this simple coupled oscillator model for explaining experimental data, with reasonably good success in fitting the locations of the eigenfrequencies~\cite{nakayama-control-2013, wang-coherent-2016, zheng-manipulating-2017, wen-room-temperature-2017, struhrenberg-strong-2018, xie-enhanced-2020, shapochkin-light-induced-2020, gomez-near-perfect-2021}. 
For example,
the coupled oscillator model was used to describe the strong coupling behavior between WSe$_2$ and a gold bipyramid MNP~\cite{struhrenberg-strong-2018}; the
work explored the coupling as a function of number of TMDC layers, which resulted in a saturation of coupling as the number of layers increased. Simulation fits with the simple model and experimental data are good, but they fail to account for the actual effective mode volume, $V_{\rm eff}$, of the MNP being a function of space, and the meaning
of $V_{\rm eff}$ is also ambiguous in general, since it is defined from an effective theoretical coupling constant, 
\begin{equation}
    g_{\rm eff} = d\sqrt{\frac{4\pi \hbar Nc}{\epsilon\epsilon_0\lambda V_c }},
    \label{eq:gth}
\end{equation}
where $d$ is the exciton dipole moment, $N$ is the number of excitons, $c$ is the speed of light in vacuum, and $\lambda$ is the wavelength; here $V_{c}$ is clearly a
{\it spatially-averaged} effective mode volume, but in reality it changes as a function of position as does $g$. This simple theory likely relates back to older works that exploit a simple coupled mode theory with translational invariance, and also without any dissipation in general, since the theory exploits the properties of
 ``normal modes'', which are not the correct cavity modes of open resonators~\cite{leung-completeness-1994,kristensen-generalized-2012}. In addition, it is not clear how to choose the ratio of number of excitons $N$ per mode volume $V_C$
Reference~\cite{lawless-influence-2020} also uses  a bipyramid nanoparticle coupled to MoSe$_2$, which employs the same assumption of a fixed and heuristic effective mode volume from a two oscillator model. Moreover, most reports in this field~\cite{wen-room-temperature-2017} do not  properly calculate the mode volume of a lossy mode
system; e.g., in this case, it was assumed that 
$V_{c} \propto (\epsilon^R +2\omega_c \epsilon^I/\gamma_c) |E_c|^2$, and $N$ is also not well defined. This is because the localized plasmons excited by the 3D nanoparticle break translational invariance, even if the TMDC-MNP hybrid system is excited with a plane wave.

A recent review paper~\cite{tserkezis-applicability-2020} gives a detailed critique on the use and warnings of using $N$ and $V_c$ in many nanophotonic systems, highlighting the limitations of Eq.~\eqref{eq:gth} and the problems that can arise. 
For example, with collective exciton systems,  usually a total coupling constant is determined as $g = \sqrt{\sum_i g(\mathbf{r}_i)^2}$, where $i$ is the $i^{\rm th}$ exciton, and the exciton Bohr radius is assumed to be constant over the entire TMDC sheet.
The authors highlighted that 
for a  TMDC coupled to plasmonic system, this is obviously not applicable, and a more appropriate calculation would include a more detailed theory of excitons as spatially extended composite particles.

To help address this problem in terms of the underlying open cavity modes, 
plasmonic cavity systems 
can be rigorously described in terms of
QNMs, which are the complex eigenmode solutions
 for resonators with open boundary
conditions. The use of QNMs
allows one to 
obtain the correct position-dependent mode volume (which simply characterizes the normalized mode strength squared) and  coupling constant as functions of space which directly involves the excitonic wave function. The same theory also gives the photonic Lamb shifts and collective effects if required, e.g., coupling to multiple
quantum dots at different spatial locations, and important propagation effects. 
These QNM  can also be fully quantized~\cite{franke-quantization-2019}
allowing for the exploration of new physics beyond the simple JC models~\cite{PhysRevResearch.2.033456}, and accounting for important QNM phase effects.
Thus, it is highly desirable to connect to the underlying 
QNMs of the MNP, before and after coupling to the TMDC system. In the presence of TMDC coupling, the ``hybrid'' QNMs fully characterize the 
classical electromagnetic coupling. 
Notably, these hybrid QNMs
are valid modes regardless of whether we work
in the weak or strong coupling regime, so they offer a unified description.
Just like two coupled atoms, which can form super-radiant and sub-radiant states, the coupling between photonic resonators can also form analogues of these hybrid states, and are the natural dressed-states of the system.

\subsection{Modelling the dielectric properties of the gold resonator and the M\lowercase{o}S\lowercase{e}$_2$ layer}
\label{sec:1}

To model the gold MNP, we employ the Drude model, 
\begin{equation}
    \epsilon_{\rm Drude}(\omega) = 1 - \frac{{\omega_p^2}}{{\omega(\omega + i\gamma_p)}},
    \label{eq:drude}
\end{equation}
 where $\omega_p$ and $\gamma_p$ are the plasma and collision frequencies, respectively.
 Although this is a local material model,
 it is known to work quantitatively well even for sub-nm gap sizes. In adidtion,
 the smallest gap sizes used below (0.5\,nm)
 are experimentally feasible and 
stay within a regime where electronic tunneling effects are
negligible~\cite{Zhu2016}.
However, note that the QNM theory we use below can also  include nonlocal effects, if needed, which has been demonstrated at the level of a hydrodynamical
model~\cite{KamandarDezfouli2017}.

To model the TMDC sheet, we begin with the 2D susceptibility for right-handed ($+$) or left-handed ($-$) polarized light, 
\begin{equation}
    {\chi}_{+/-}^{\rm 2D}(\omega) = \frac{\vert \mathbf{d}^{\rm 1s}_{\rm +/-}\vert^2}{\hbar\epsilon_0}
    \left( \frac{1}{\omega_0^{\rm 1s} - \omega-i\gamma_{\rm 1s}^\prime  } \right),
    \label{eq:chi2d}
\end{equation}
where $\epsilon_0$ is vacuum permittivity,
and $\rm \mathbf{d}^{1s}_{+/-}=d\,\varphi^{1s}(\mathbf{r} = 0)\,\mathbf{e}_{+/-}$ is the excitonic dipole moment, exhibiting a circular dichroism. The probability of finding electrons and holes at the same position $\varphi^{1s} (\mathbf{r} = 0)$  is calculated by exploiting the Wannier equation~\cite{haug2009quantum,kira2006many} where we use the reduced excitonic mass~\cite{kormanyos2015k} and the Coulomb potential of the slab
~\cite{trolle2017model} as an input. $\gamma'_{\rm 1s}$ 
is the dephasing rate from exciton phonon coupling, which was calculated according to Ref.~\cite{selig2016excitonic} (note that this rate does not include radiative decay as that will be self-contently captured from a self-consistent Maxwell model that we use below).

As only small excitonic momenta on the order of the inverse nanorod radius contribute to the interaction it is reasonable to assume a flat excitonic dispersion in order to reduce numerical complexity. For our parameter choice in MoSe$_2$, this approximation overestimates the Rabi splitting by a few meV \cite{selig2016excitonic} which is small compared to the observed splitting.

Since the unit of $\chi^{2D}$ is m (length), 
using Eq.~\eqref{eq:chi2d}, we obtain the 
unitless 3D electric permittivity (or dielectric constant),
\begin{equation}
    {\epsilon}_{+/-}^{\rm 3D}(\omega) = {1} + \frac{{\chi}_{+/-}^{\rm 2D}(\omega)}{b},
    \label{eq:eps3d}
\end{equation}
where $b$ is the thickness of the TMDC layer. 

\begin{figure}[htb]
    \centering
    \includegraphics[trim=0cm 0cm 0cm 0cm, clip=true,width=1\columnwidth]{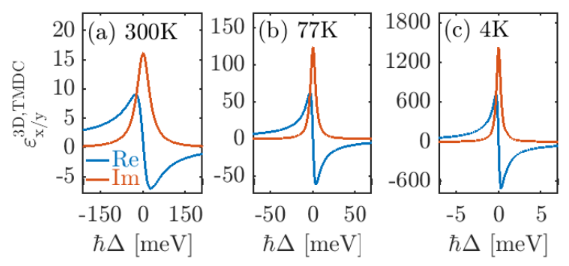}
    \caption{(a-c) The relative in-plane permittivity of monolayer MoSe$_2$ given by Eq.~\eqref{eq:epsmatrix} as a function of detuning from the TMDC resonance frequency, $\rm \hbar\Delta = \hbar(\omega - \omega_0^{1s})$, for selected temperatures (a-c).  Parameters are given in Tab.~\ref{tab:parameters}. }
    \label{fig:permittivity3d}
\end{figure}

All of the parameters pertaining to the 1s exciton for MoSe$_2$ encapsulated in hBN are given in Tab.~\ref{tab:parameters}; note that there is also evidence that encapsulating TMDC layers in hBN is helpful in reducing defects as well as reducing the linewidth of the excitonic resonance \cite{martin-encapsulation-2020}. The low-temperature decay rate in Tab.~\ref{tab:parameters} agrees with the results in Ref.~\onlinecite{martin_encapsulation_2020}. The permittivity of the TMDC, as a function of frequency, is shown in Fig.~\ref{fig:permittivity3d} for $T = 4, 77$ and $300$~K.

\begin{table}[h!]
	\centering
	\caption{Material parameters for MoSe$_2$ encapsulated in hBN, as well as gold. }
	\begin{tabular}{c c c}
	    Parameter & Value & Reference \\ 
 		\hline
		$\rm d\phantom{\int_a^b}$ &    4.3254e-29~Cm ($\sim$ 13 D) & \onlinecite{xiao2012coupled} \\
		$\rm \varphi^{1s}(\mathbf{r} = 0)$ & 4.8e8~m$^{-1}$  & $^*$\\
		$\phantom{\int_a^b}\rm \hbar \omega_0^{1s}$ & 1.6~eV &\onlinecite{christiansen2017phonon}\\
		$\phantom{\int_a^b}\rm \hbar \gamma'_{1s}(4~K)$ & 0.3~meV &$^{**}$\\
		$\phantom{\int_a^b}\rm \hbar \gamma'_{1s}(77~K)$ & 3.5~meV &$^{**}$\\ 
		$\phantom{\int_a^b}\rm \hbar \gamma'_{1s}(300~K)$ & 26.9~meV  &$^{**}$\\
		$b$ & 0.7~nm &\\
		$\phantom{\int_a^b}\rm \frac{({d} \varphi^{1s})^2}{\epsilon_0 b\phantom{\int_a^b}}$  & 0.434 eV &\\
		\hline
		$\phantom{\int_a^b}\rm \varepsilon_{hBN}$ & 4.5 & \onlinecite{brem2019intrinsic},\onlinecite{geick1966normal}\\
		\hline
		$\phantom{\int_a^b}\rm \hbar \omega_p$ & 8.2935~eV & $^{***}$\\
		$\rm \hbar \gamma_p$ & 9.3~meV&$^{***}$\\
        \hline	
	\end{tabular}
	\label{tab:parameters}
	
{\footnotesize{
$\ \ ^*$ calculated by exploiting the Wannier equation

$\ ^{**}$ calculated by exploiting the method from ~\cite{selig2016excitonic}

$^{***}$ parameters used to fit Palik data for gold \cite{palikgold}
}}\\

\end{table}

In a 2D system (i.e., a sheet), the permittivity is naturally anisotropic. In the circular basis, the 2D susceptibility tensor is written as
\begin{equation}
    {\bm \chi}^{\rm 2D}_{\rm+/-} \rightarrow 
    \begin{bmatrix} 
    \chi_+^{\rm 2D} & 0 \\ 0 & \chi_-^{\rm 2D}      
    \end{bmatrix},
\end{equation}
where the $\omega$ dependence is implicit. This can be transformed into the Cartesian basis such that $\mathbf{e}_{+/-} = \frac{1}{\sqrt{2}}\left(\mathbf{e}_x \pm i\mathbf{e}_y\right)$; thus we  write
\begin{equation}
\rm
   {\bm  \chi}_{xy}^{2D} \rightarrow 
    \frac{1}{2}\begin{bmatrix} 
    \chi_+^{\rm 2D}+ \chi_-^{\rm 2D} & -i\left(\chi_+^{\rm 2D} - \chi_-^{\rm 2D} \right) \\ 
    i\left(\chi_+^{\rm 2D} - \chi_-^{\rm 2D} \right) & \chi_+^{\rm 2D} + \chi_-^{\rm 2D}
    \end{bmatrix},
\end{equation}
which for the case of equivalent valleys ($\chi^{\rm 2D}_+ = \chi^{\rm 2D}_-$), reduces to
\begin{equation}
    {\bm \chi}_{xy}^{\rm 2D} \rightarrow 
    \begin{bmatrix} 
    \chi_x^{\rm 2D} & 0 \\ 
    0 &  \chi_y^{\rm 2D}
    \end{bmatrix},
\end{equation}
where $\chi_{x}^{\rm 2D}=\chi_{y}^{\rm 2D}=\chi_{+/-}^{\rm 2D}$. The full 3D permittivity tensor in Cartesian coordinates is then given by
\begin{equation}
    {\bm \epsilon}_{xyz}^{\rm 3D} \rightarrow 
    \begin{bmatrix} 
    1 + \chi_x^{\rm 2D}/b & 0 & 0 \\ 
    0 & 1 + \chi_y^{\rm 2D}/b & 0 \\
    0 & 0 & \epsilon_{\rm hBN}
    \end{bmatrix},
    \label{eq:epsmatrix}
\end{equation}
for which the $x$ and $y$ components are plotted for three different temperatures in Fig.~\ref{fig:permittivity3d}.

\section{Metallic resonator designs and electromagnetic calculations of strong coupling and figures of merit}

\begin{figure}[t!]
\includegraphics[width=1\columnwidth]{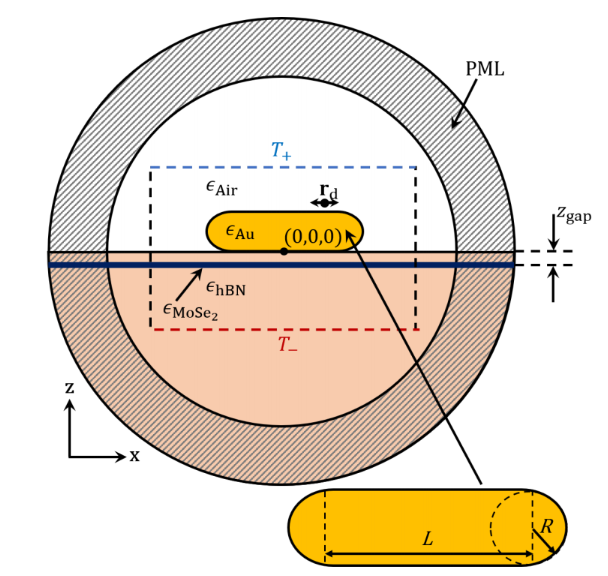}
\caption{Schematic of the numerical simulation geometry, containing a gold nanorod on top of a hBN-MoSe$_2$-hBN stack, where $z_{\rm gap}$ is the distance between the nanorod and the top of the TMDC, the dipole location is given by $\mathbf{r}_{\rm d}$, and the top(bottom) power transmission monitors, $T_+(T_-)$, are shown as dashed blue(red) lines. The transmission monitors are defined by a cylindrical surface with height of 180~nm and radius of 120~nm, centered at the origin.  The entire simulation domain is surrounded by perfectly matched layers (PMLs) to simulate outgoing boundary conditions.}
\label{fig:fullscheme}
\end{figure}

For the MNP, we  use a simple cylindrical gold nanorod with rounded ends (see Fig.~\ref{fig:fullscheme}) where the radius (R) and length (L) can be tuned to achieve a spectral resonance that overlaps with the TMDC resonance. The second lowest mode (with respect to frequency) of the MNP is chosen to couple to the TMDC as it  is typically much larger in quality factor (thus, smaller decay rate), compared to the first order mode, which can aid in obtaining strong coupling behaviour with the TMDC; note that Ref.~\cite{zheng-manipulating-2017} uses the third lowest mode of a silver nanorod to achieve high-$Q$ low-mode-volume conditions, requiring a much larger MNP. 
All calculations discussed below
are fully 3D in nature without any approximations or fitting parameters,
other than the microscopic parameters already stated in the models. 



For our models, we will first compute the
MNP QNM as well as the hybrid QNMs formed from the coupled TMDC-MNP system, at different gap separations and for different temperatures. The QNMs are spatially dependent and contain important phase effects and frequency information (e.g., from the complex poles).
Using the QNMs in a Green function expansion,  we can easily compute a range of important light-matter interactions at any spatial position over a wide range of frequencies, and analytically. We also stress again, that these QNMs can also be fully quantized for use in a rigorous open-system quantum optics model~\cite{franke-quantization-2019} if needed.
In typical experiments, the absorption, transmission, or reflection from an optical source is usually what can be used to detect this mode hybridization, e.g., in combination with near field optical techniques. One can also add additional emitters such
as single quantum dots, and study the modified spontaneous emission (Purcell effect), as well as light propagation and photon exchange effects.

As an application of the theory, we will report on two frequency-dependent functions that are relevant to probing strong coupling. The first is the Purcell factor (PF) as a function of frequency, which is related to the modified spontaneous emission of a point dipole at some position ${\bf r}_d$. The Purcell factor depends on the projected local density of states (LDOS), which is completely contained within the imaginary part of the (photonic) Green function, ${\rm Im}\left[{\bf G}({\bf r}_d,{\bf r}_d,\omega)\right]$. Since we are interested in the strong coupling between the TMDC and the MNP, we consider an immutable dipole (weakly coupled and in a linear response, though this is not a model restriction).
Second, we consider the emitted 
spectrum at some surface away from a dipole source, which can be probed, e.g., by 
near field microscopy techniques and localized detectors. For this latter case, we require the two-space-point Green function, ${\bf G}({\bf r},{\bf r}_d,\omega)$, projected
on a surface point ${\bf r}$ at a different position away from the dipole. This scenario differs in the sense that it also accounts for quenching and propagation between two spatial points, unlike the LDOS. These spectral function also relate to the QNM phase (at different positions), and are complimentary but subtly different. In a traditional normal mode approach, these spectral function would typically be identical, unless additional filtering is included in the input-output theory. 

To numerically implement and fully corroborate our QNM theory and model assumptions, we  carry out full
numerical electromagnetic simulations in COMSOL Multiphysics \cite{comsol}. The numerical generalized PF (denoted as $ F_{\rm P}$), 
which includes the homogeneous background contribution of $\epsilon_{\rm B}=1$ as well the scattered contributions of all of the modes present in the nanostructure, is calculated by performing full dipole simulations and examining surface-integrated Poynting vector around the dipole at $\mathbf{r}_{\rm d}$ (cf.~Fig.~\ref{fig:fullscheme}), from
\begin{equation}
    F_{\rm P}(\mathbf{r}_{\rm d},\omega) = \frac{\int_s \hat{\mathbf{n}}\cdot\mathbf{S}_{\rm dipole_i, total}(\mathbf{r};\omega) \rm{d}A}{\int_{s} \hat{\mathbf{n}}\cdot\mathbf{S}_{\rm dipole_i, background}(\mathbf{r};\omega) \rm{d}A}, 
    \label{eq:comsolFP}
\end{equation}
where the surface $s$ is of a small sphere centred around the {\it finite-size} dipole oriented in the $i^{\rm th}$
direction; the small sphere is approximately 1~nm in radius, and $\hat{\bf{n}}$ is the normal vector directed outward relative to the dipole.
This allows us to capture the power flow which we have also checked is quantitatively accurate against known solutions in free space. In experiments, this would be measured as the ratio of the spontaneous emission rate of a dipole emitter in the full scattering structure to the rate in a homogeneous background medium.
COMSOL can also be used to obtain the QNMs of the system which allows us to obtain the PF everywhere in the system, not just at a single point as in Eq.~\eqref{eq:comsolFP}.

The gold nanorod (on the hBN substrate) was designed such that the peak PF of the second lowest mode
was located near 1.6~eV to spectrally overlap with the TMDC response.
The final dimensions of the nanorod were determined to be $L = 148$~nm and $R = 10$~nm. The small frequency shift due to the presence of the TMDC, which alters the effective background permittivity, was not considered in this design, but will be discussed later when examining the anti-crossing behavior with respect to detuning.

\section{Hybrid quasinormal modes and spectral splitting in the strong coupling regime} 

\begin{figure*}[htp]
    \centering
    \includegraphics[width=0.95\textwidth]{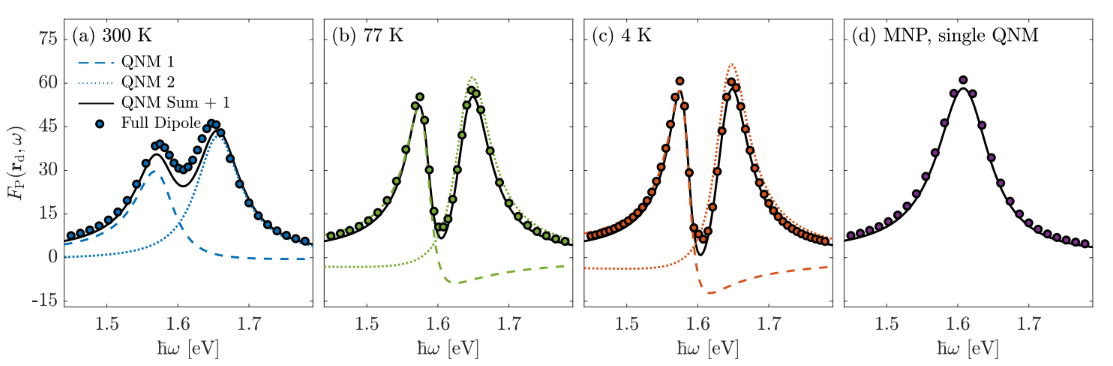}
     \caption{
     Generalized Purcell factor at the $x$-oriented dipole location ($\rm \mathbf{r}_d$~=~[44,~0,~40]~nm) using full dipole calculations (markers, Eq.~\eqref{eq:comsolFP}) and QNM analysis (lines, Eq.~\eqref{eq:qnmPF}). Generalized Purcell factor for (a-c) the hybridized TMDC-MNP modes for selected temperature and (d) the MNP alone on the hBN substrate. $L$~=~148~nm, $R$~=~10~nm, $ z_{\rm gap}=~2$\,nm.  Note that some of the full-dipole PF features near the minimum are caused by finite size effects, and small reflections that can come from the TMDC interfering with the PML; thus, this region shows a larger departure with the QNM results, but the overall trends are clearly in very good qualitative agreement, especially as we are not using any fitting parameters. }
    \label{fig:temp}
%
    \includegraphics[width=0.95\textwidth]{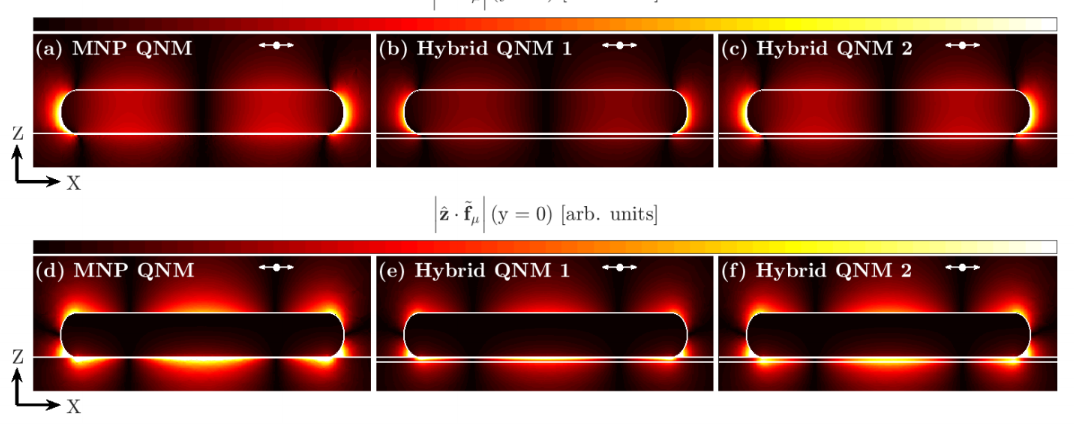}
    \caption{ The magnitude of the (a-c) $x$-component and (d-f) $z$-component of the QNM field profile for the original MNP QNM, first QNM of the hybrid MNP-TMDC system, and second QNM of the hybrid MNP-TMDC system at a temperature of 4~K. $L=148$~nm, $R=10$~nm, $z_{\rm gap}=1$ nm. White lines outline the MNP, substrate, and TMDC. The dipole location is given by the double arrow and marker, which is located 20~nm above the gold surface. }
    \label{fig:absF}
\end{figure*}
\begin{figure*}[htp]
    \centering
    \includegraphics[width=\textwidth]{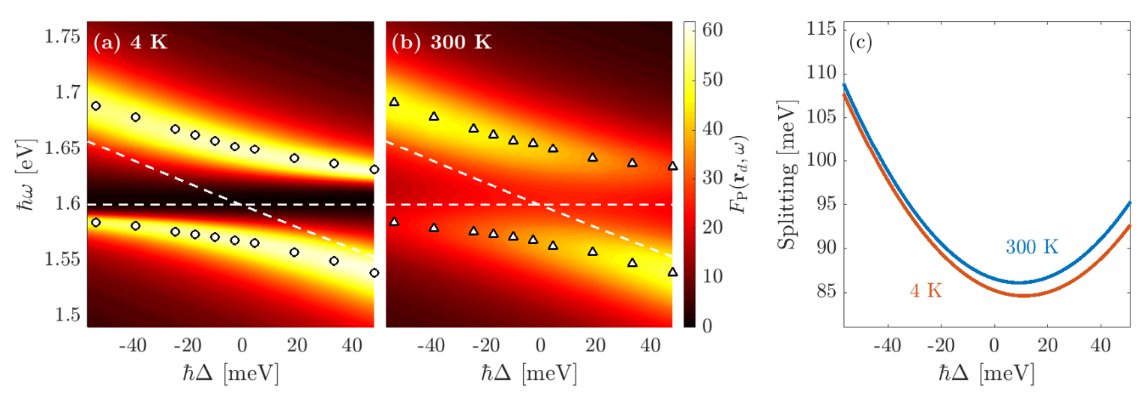}
    \caption{Generalized Purcell factor at the location of the $x$-oriented dipole ($\mathbf{r}_d $~=~[44,~0,~40]~nm) using QNM analysis (Eq.~\eqref{eq:qnmPF}) as a function of detunining ($\rm \Delta \equiv \omega_0^{1s} - \omega_0^{MNP}$) at (a) 4~K and (b) 300~K. The dashed lines show the resonant frequency of the TMDC and the MNP. Detuning is achieved by changing the length of the MNP ($L=140-156~$nm). The markers indicate the peak PF of the upper and lower peaks at the values of detuning used to create the image. (c) The splitting of the upper and lower peaks at each temperature where the line is a parabolic fit to the data. Minimum anti-crossing is observed at $\approx 10$ meV detuning (see text). }
    \label{fig:detune}
\end{figure*}
\begin{figure}[htp]
    \centering
    \includegraphics[width=\columnwidth]{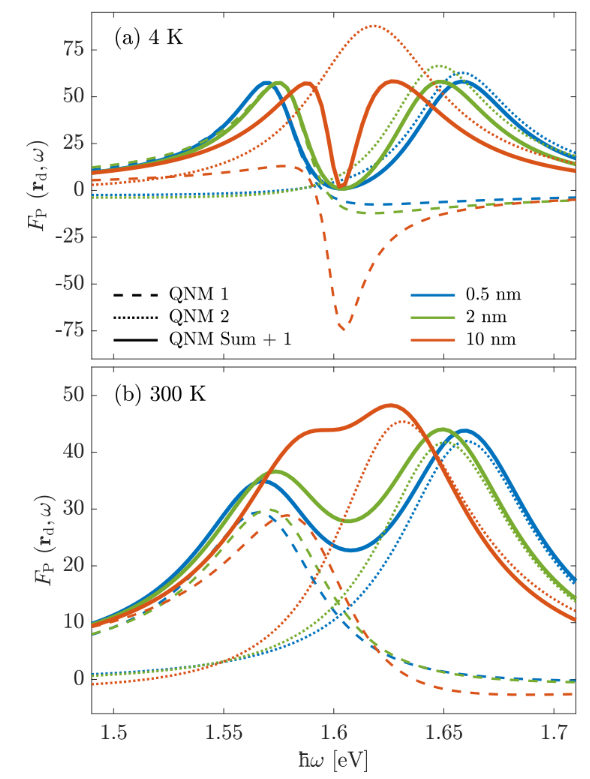}
    \caption{Generalized Purcell factor at the location of the $x$-oriented dipole ($\rm \mathbf{r}_d $~=~[44,~0,~40]~nm) using QNM analysis (Eq.~\eqref{eq:qnmPF}) for the individual QNMs of the hybridized system (dashed and dotted lines) as well as the sum of the QNMs (solid lines), as a function of temperature and gap size. }
    \label{fig:zgap}
\end{figure}

We next describe 
how the hybrid QNMs can be obtained 
to assess and understand the strong coupling regime of the coupled TMDC MNP system and the spectral interaction of QNMs, i.~e., through ``QNM splitting''. Numerically, we again use COMSOL, since  it can 
work in complex frequency space.
We adopt an inverse Green function approach,
where the normalized QNM are obtained from
the solution to a scattering problem
in complex frequency space~\cite{bai-efficient-2013,carlson-dissipative-2020}.

The optical QNMs are solutions to the 
vector 
Helmholtz equation in the complex frequency domain, which are the proper mode solutions of cavity structures with open-boundary conditions. Assuming non-magnetic ($\mu_r = 1$) permeability, the electric field QNM eigenvalue equation is
\begin{equation}
\boldsymbol{\nabla}\times\boldsymbol{\nabla}\times\tilde{\mathbf{f}}_{{\mu}}\left(\mathbf{r}\right)-\left(\dfrac{\tilde{\omega}_{{\mu}}}{c}\right)^{2}
{\bm \epsilon}\left(\mathbf{r},\tilde{\omega}_\mu\right)\,\tilde{\mathbf{f}}_{{\mu}}\left(\mathbf{r}\right)=0,
\label{eq:helmholtz}
\end{equation}
where $\tilde{\mathbf{f}}_{{\mu}}$ are the QNMs, $\tilde{\omega}_\mu = \omega_\mu - i\gamma_\mu$  are the complex eigenfrequencies, and $\mu$ is the mode number.
The permittivity, $\mathbf{\epsilon}(\mathbf{r},\omega)$, is completely general and is composed of the Drude gold (Eq.~\eqref{eq:drude}), the anisotropic TMDC model (Eq.~\eqref{eq:epsmatrix}), the background permittivity ($\epsilon_{\rm B}$), and hBN substrate permittivity ($\epsilon_{\rm hBN}$). The QNM cavity factors are given
by $Q_\mu = \omega_\mu/2\gamma_\mu$.

The electric-field Green function
can then be defined in terms of an expansion over the QNMs and a complex frequency prefactor~\cite{kristensen-modeling-2020},
%
\begin{equation}
\mathbf{G}(\mathbf{r}, \mathbf{r}^\prime, \omega) = \sum_{\mu=\pm1,\pm2, \cdots}
\frac{\omega}{2(\tilde{\omega}_\mu - \omega)}
\tilde{\mathbf{f}}_{\mu}
(\mathbf{r}) \tilde{\mathbf{f}}_\mu (\mathbf{r}^\prime),  
\label{eq:2}
\end{equation}
where
the QNMs only depend on spatial position. Note that the form of the Green function uses an unconjugated product, and   this
is a general feature of QNMs, including the MNP on its own as well as the hybrid modes.
The unconjugated product
 is essential to capture complex interference effects arising from the QNM phase of the hybridized modes, which can also gives rise to Fano-like resonances~\cite{RosenkrantzdeLasson2015,2017PRA-hybrid}.
The above expansion is known to be highly
accurate for spatial locations within and near the resonator~\cite{kristensen-modes-2014,lalanne-light-2018,kristensen-generalized-2012}. 
For positions far from the resonator, then the above solution can be used with the Dyson equation to obtain regularized (non-divergent) QNMs~\cite{ge-quasinormal-2014}, even for spatial positions far from the resonator; or one can also use
near-field to far-field transformations~\cite{PhysRevB.101.205402} (see also Ref.~\onlinecite{PhysRevB.102.035432}).

As mentioned earlier,  the concept of effective mode volume here  represents a measure of $\tilde{\bf f}_\mu^2$, and is thus  a
spatially dependent quantity, and it is also complex (for QNMs).
The complex mode volume, for mode $\mu$, 
for an emitter at position ${\bf r}_{\rm d}$
in $\epsilon_{\rm B}$,
is given by
\begin{equation}
    \begin{aligned}
      \tilde{V}_\mu(\mathbf{r}_{\rm d}) = \frac{1}{\epsilon_{\rm B}(\mathbf{r}_{\rm d})\tilde{\mathbf{f}}_\mu(\mathbf{r}_{\rm d})^2}.
    \end{aligned}
    \label{eq:v}
\end{equation}

The effective mode volume for use in Purcell's formula can be calculated as the real part of Eq.~\eqref{eq:v},
\begin{equation}
    \begin{aligned}
      {V}_{\rm eff,\mu}(\mathbf{r}_{\rm d}) = \rm Re[\tilde{V}_\mu(\mathbf{r}_{\rm d})].
    \end{aligned}
    \label{eq:veff}
\end{equation}
Thus, for one QNM, $\mu=c$ (positive frequency pole), we recover the usual one mode solution from
the Purcell factor,
\begin{equation}
{\rm PF}({\bf r}_{\rm d}, \omega=\omega_c)
= \frac{3}{4\pi^2}\left (  \frac{\lambda_c}{n_{\rm B}}  \right  )^3
\frac{Q_c}{V_{{\rm eff},c}({\bf r}_{\rm d})},
\end{equation}
where we assume the dipole is aligned with the dominant polarization of the QNM, and $n_{\rm B}=\sqrt{\epsilon_{\rm B}}$. With several modes,  there are complex interference
effects stemming from the concept of a complex effective mode volume, which is a consequence of the QNM phase.
These effects can also be obtained directly from the single QNM Green function, which is in fact much more convenient with coupled QNMs.

For several QNMs,  
the generalized PF can be obtained from the two QNM Green function ($\mu=1,2$) relative to the homogeneous solution,
\begin{equation}
    {F}_{\rm P}(\mathbf{r}_{\rm d},\omega) = \frac{\hat{\mathbf{n}}_i\cdot {\rm Im}[\mathbf{G}(\mathbf{r}_{\rm d},\mathbf{r}_{\rm d},\omega)]\cdot\hat{\mathbf{n}}_i}{{\rm Im}[G_{\rm hom}]} +1,
    \label{eq:qnmPF}
\end{equation}
where ${\rm Im}[G_{\rm hom}] = \frac{n_{\rm B}}{6\pi\epsilon_0}\left(\frac{\omega}{c}\right)^3$.
The plus one appears due to the total Green function being the sum of the scattered and the homogeneous parts~\cite{GeNJP2014}.

Figure~\ref{fig:temp} shows the numerical full-dipole PF (markers) as well as the individual hybrid QNMs (dashed/dotted lines) and the sum of the two QNMs (solid line) for the hybrid TMDC-MNP system at various temperatures of interest (4~K, 77~K, 300~K) as well as the single MNP on hBN (independent of temperature). Considering no other modes are used in the sum (i.e., higher and lower order plasmonic modes that are not spectrally isolated from the frequency range of interest), 
the agreement between the full numerical simulations and the two QNM simulations (with no fitting parameters) is overall in very good agreement.

Having now confirmed that the dipole PF is well described by the two QNM
Green function expansion, we stress that  Eq.~\eqref{eq:2} can be used to obtain the PF at all locations near the resonator, highlighting the remarkable power of the QNM technique. This can be recognized from Figure~\ref{fig:absF} showing the absolute value of the $x$-component of the individual QNMs, $\vert \hat{\mathbf{x}}\cdot \tilde{\mathbf{f}}_\mu(\mathbf{r}) \vert$, in the $xz$-plane ($y=0$) and $xy$-plane ($z=-0.5$~nm, in the middle of the gap), for a gap of 1~nm. The MNP, substrate, and TMDC are outlined as white lines. The two hybridized modes look qualitatively similar to the bare MNP QNM, but are indeed new modes, not just perturbations of the original.   
Figure \ref{fig:detune} shows the anti-crossing behavior in the total PF, calculated using Eq.~\eqref{eq:qnmPF}, for the MNP-TMDC system with a gap of 2~nm which is expected in the presence of strong coupling, showing a minimum Rabi splitting of 84.6~meV and 86.1~meV
 for temperatures of 4~K and 300~K, respectively. Panel (c) shows the value of the Rabi splitting as a function of detuning, which is achieved by modifying the length of the MNP. In addition to the anti-crossing behavior, $\rm \hbar\Omega_R > \hbar(\gamma_{p} + \gamma_{ex})$ is sometimes used as a benchmark for the onset of strong coupling; in this case, these values are 45.1~meV and 71.7~meV, respectively, which are much less than the observed splitting.  
 
 The minimum Rabi splitting is observed at a detuning of approximately 10-15~meV, rather than 0~meV. This is due to finite QNM losses as well as the procedure used to design the MNP; the radius and length of the MNP (on a substrate of hBN) were varied until the peak of the PF matched the resonance frequency of the TMDC. However, due to the presence of the TMDC in the system, it slightly shifts the background effective permittivity which blue shifts the MNP resonance. The peak permittivity is larger for lower temperature, explaining why the shift is slightly larger at 4~K relative to 300~K. A 10~meV shift can be achieved by changing the effective substrate permittivity from 4.5 to 4.6, which is reasonable magnitude to expect from the presence of a TMDC layer.  
 
The effect of the gap size between the TMDC and MNP, $z_{\rm gap}$, is shown in Fig.~\ref{fig:zgap} at 4~K and 300~K. The individual QNM PFs are shown as dashed/dotted lines and the total PFs are shown as solid lines. As expected, the observed Rabi spitting increases as the TMDC gets closer to the MNP since the strength of the electric field due to the plasmonic resonance increases dramatically near the metal surface. Interestingly, the Fano-like behavior of the individual QNMs is most pronounced at either large gap sizes or low temperatures. This can be explained by the QNM phase interference between the bare modes, which is more drastic with the inequality of the linewidths of the two modes,  and is more pronounced 
at elevated temperatures.

\begin{figure}[h]
    \centering
    \includegraphics[width=\columnwidth]{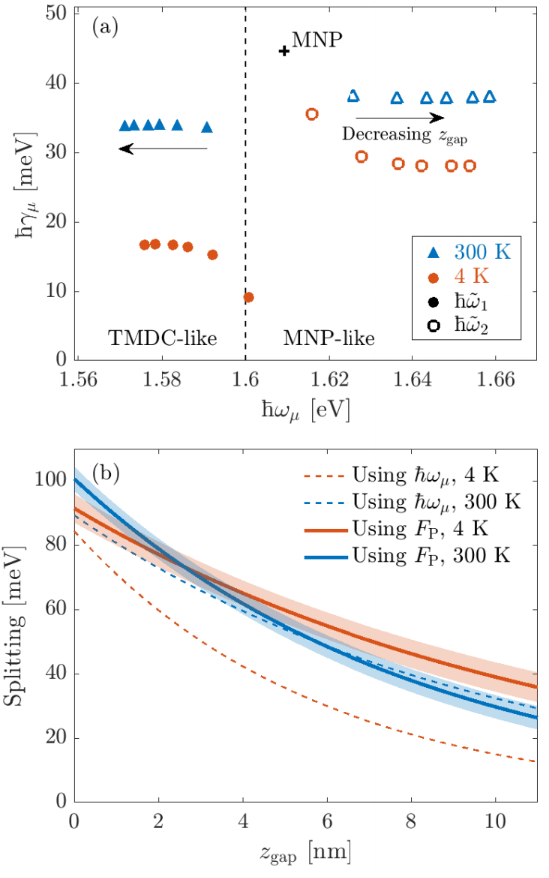}
    \caption{
    (a) 
    Complex poles of the QNMs for the hybridized system at 4~K and 300~K as a function of gap size (values of 0.5, 1, 2, 3, 5, and 10~nm). The TMDC does not support a QNM, so the resonant frequency ($\rm \omega^{1s}$) is given by a dashed line. The complex pole of the MNP on the hBN substrate is given by the plus (+) marker. (b) Splitting calculated using the real part of the complex poles (dashed lines) and using the frequency at the maxima of the generalized Purcell factor (solid lines) for 4~K (red) and 300~K (blue). These lines are exponential fits to the data shown in (a), and the standard deviation of the Purcell splitting is shown by the shaded region; the standard deviation of the pole splitting is too small to see on this scale.}
    \label{fig:poles}
\end{figure}

\section{Complex poles as a function of temperature and gap size}

Next,  Fig.~\ref{fig:poles}(a) displays a summary of the complex poles of the hybrid QNMs as a function of temperature and gap size. Also shown is the resonance frequency of the TMDC alone (dashed) and the QNM pole of the MNP alone (marker, `$+$'). Since the TMDC alone does not support a QNM, the TMDC resonance is represented by a vertical line. As the gap between the TMDC and MNP increases, the modes become less coupled, resulting in the modes moving towards either the MNP (upper mode) or TMDC (lower mode), until they are completely uncoupled. The upper mode is more MNP-like, while the lower mode is a more pronounced hybridization of the two original modes. The difference in the imaginary part of the modes contributes to greater changes in phase which in-turn result in greater Fano-like lineshapes seen in Fig.~\ref{fig:zgap}. In panel Fig.~\ref{fig:poles}(b), the Rabi splitting as a function of gap size for 4~K and 300~K is shown using two different calculations: (i) the splitting due to the poles (dashed line) is calculated as the difference between the real part of the poles; and (ii) the splitting due to the PF (solid line) is calculated as the difference between the frequency locations of the two maximums of the total PF. Since the data is obtained for discrete values of $\rm z_{\rm gap}$, the data is fit to an exponential curve to make the readability of the information more clear. The standard deviations of the fits are shown by the shaded region, which is negligible for the pole-calculated fit. We can see that, overall, the splitting is larger when observed from the PF than from the poles. 

\section{Dipole transmitted power and the Poynting vector using the QNMs, and spectral splitting from the Purcell factor versus the transmitted power}
\label{sec:trans}

\begin{figure}[htp]
    \centering
    \includegraphics[width=0.99\columnwidth]{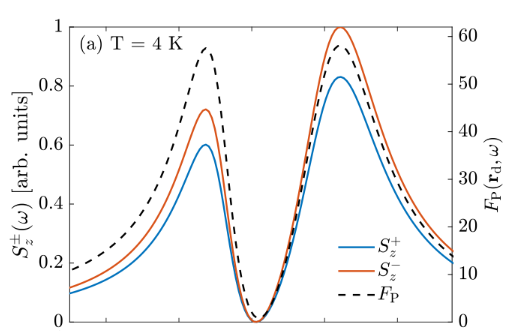}
    \includegraphics[width=0.99\columnwidth]{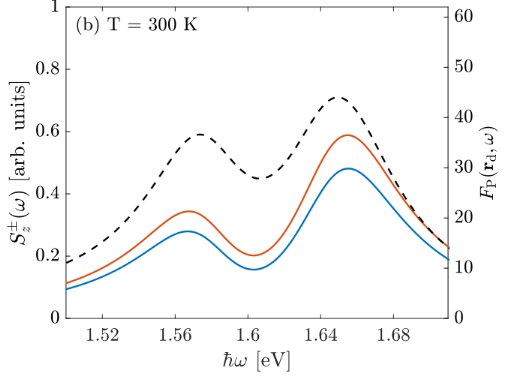}
    \caption{Transmission and reflection of the Poynting vector, reconstructed using the QNM analysis. $z_{\rm gap}$=2~nm and ${\bf r}_{\rm d} = [44,~0,~40]$~nm at (a) 4~K and (b) 300~K. }
    \label{fig:TR}
\end{figure}

As discussed earlier, another experimental observable could be the emitted spectrum to some detector position that is away from 
a local excitation (such as from a dipole source). The theory to describe the detected field at these points requires the full two-space-point Green function (or {\it propagator}), rather than just the projected LDOS.
In normal mode theory, such a spectrum would be identical to the one at the dipole location (apart from a scalign constant), but in QNM theory the phase of the QNM changes as a function of position. In addition, there are propagation and quenching effects between two spatial position, than can render these two function different in plasmonic and TMDC systems.
Thus,  for completeness, here we also describe how the Poynting vector (and thus, the dipole transmission through some arbitrary surface) can be reconstructed using the QNMs in the relevant spectral region of interest (in a spatial region that is not too far from the resonators).

The Poynting vector is defined as,
\begin{equation}
\mathbf{S}(\mathbf{r},\omega) = \frac{1}{2}{\rm Re}\left[\mathbf{E}(\mathbf{r},\omega)\times\mathbf{H}^*(\mathbf{r},\omega)\right],
\label{eq:s}
\end{equation}
where $\mathbf{E}$ is the electric field, $\mathbf{H}$ is the auxillary magnetic field ($\mathbf{H} = \mu_0^{-1}\mathbf{B}$, where $\mu_0$ is the permeability of free space and $\mathbf{B} = (i\omega)^{-1}\boldsymbol{\nabla}\times\mathbf{E}$ is the magnetic field). The electric field due to a point dipole source with dipole moment ${\mathbf{d}}$ is constructed from the Green function via,
\begin{equation}
 \mathbf{E}(\mathbf{r},\omega) = \frac{1}{\epsilon_0} {\mathbf{G}(\mathbf{r},\mathbf{r}_{\rm d},\omega)}\cdot {\mathbf{d}},   
\end{equation}
where we note that the Green function is a tensor of rank 2 (3 by 3 matrix).  The total power flow through a surface $T$ is then given by, 
\begin{equation}
    S_z^\pm(\omega) = \int_{T^\pm} {S}_z(\mathbf{r},\omega)~{\rm d}\mathbf{r}.
\end{equation}

The upper surface ($T^+$) and lower surface ($T^-$)
values are shown in 
 Fig.~\ref{fig:fullscheme}, which are monitors in the $xy$ plane with normal vectors pointing outwards. The observed spectral splitting due to the power transmission of a dipole for a gap of 2~nm and temperature of 4~K is 74.1~meV. In contrast, the splitting observed in Fig.~\ref{fig:poles} due to the complex poles and the PF were 59.5~meV and 73.8~meV, respectively. There is a slightly greater splitting when observed in the transmission of a dipole emitter compared to the Purcell enhancement of the emission of the same dipole emitter.
 This is a consequence of the changing QNM from the dipole location to the surfaces for the power flow.

 Power transmission can typically be measured experimentally by exciting the system with a plane wave or tightly focussed beam \cite{sendur-near-2009} and comparing the optical transmission with and without the optical scatterer. This is different from what we have presented here, since our source is a dipole emitter, but the transmission should be similar to a  background free measurement. 
 In contrast, 
 experimentally determining the PF requires measuring the change in the radiative lifetime for a quantum emitter of either the photoluminescence spectra \cite{noda-spontaneous-2007} or Raman spectra \cite{checoury-deterministic-2010}. These differences
 are subtle but fully captured in our model.
The two-space-point
Green functions can also be used to compliment other formalism for systems containing two and more quantum emitters,
since they account for important photon exchange effects~\cite{PhysRevB.92.205420,Feist2020}




\section{Conclusions}

In summary, we have 
presented a QNM approach for the description of the modal physics of a hybrid TMDC-MNP system in the strong coupling regime. 
In this fully 3D approach, there is no fixed $g$, but rather the splitting behaviour is captured from the underlying hybrid modes, using the full 3D geometry of hybrid cavity structure.

We also examined the effect of temperature, gap size, and detuning to illustrate the characteristics of the coupling between the TMDC and MNP. Spectral splittings as large as 90~meV are observed for a gap size of 0.5~nm. Fano-like contributions from the individual QNMs also manifest in the splitting associated with strong coupling. These hybrid modes
are indeed the desired dressed-states states of the hybrid system. Knowing the QNMs of the system allows us to examine the coupling as a function of frequency and space, which can be used to examine more exotic and complex systems such as coupled arrays of dipoles or quantum emitters. Moreover, the use of a QNM model is critically important for the quantization of these lossy modes in quantum optics since they contain the full dispersive nature of the modes that arises from complex eigenfrequencies, unlike traditional normal modes that require phenomenologically added dephasing contributions in Jaynes-Cummings-type models \cite{franke-quantization-2019,PhysRevResearch.2.033456}.
Our approach thus paves the way to describe such complex cavity structures, including the coupling of more quantum emitters, fully accounting for
collective effects, as well as radiative and nonradiative decay and transport as mediated from the hybrid modes.\\


{\it Note added}: during completion of this paper, 
we became aware of  related work on the coupling
between TMDCs and MNPs~\cite{denning-quantum-2021}. Similar to our work, they consider a MNP cavity that is described by a single QNM which is then coupled to a TMDC with in-plane Coulomb interaction coupled exciton states. The coupling to the 
TMDC sheets is then carried out using 
a quantum reaction-coordinate approach
to exciton-resonator interactions, and master equations, as well as semiclassical 
calculations for the scattered field using a
Lippmann-Schwinger equation. Unlike our approach which is based on a microscopic material theory to calculate the underlying {\it hybrid  QNMs} for linear optics (and also for use 
for quantized QNM master equations~\cite{franke-quantization-2019,PhysRevResearch.2.033456}), Ref.~\cite{denning-quantum-2021} develops a coupled oscillator model with microscopically determined parameters that can be used to also describe a non-linear optical response. 

\acknowledgements
We acknowledge fruitful discussions with Dominik Christiansen and Florian Katsch (TU Berlin). We  acknowledge funding from Queen's University,
the Canadian Foundation for Innovation, 
the Natural Sciences and Engineering Research Council of Canada, and CMC Microsystems for the provision of COMSOL Multiphysics.
We also acknowledge support from the Deutsche Forschungsgemeinschaft (DFG) through SFB 951 Project B12 (Project number 182087777), Project SE 3098/1 (Project number 432266622) and the 
Alexander von Humboldt Foundation through a Humboldt Research Award.
This project has also received
funding  from  the  European  Unions  Horizon  2020
research and innovation program under Grant Agreement
No. 734690 (SONAR).

\bibliography{main.bib}

\begin{thebibliography}{72}%
\makeatletter
\providecommand \@ifxundefined [1]{%
 \@ifx{#1\undefined}
}%
\providecommand \@ifnum [1]{%
 \ifnum #1\expandafter \@firstoftwo
 \else \expandafter \@secondoftwo
 \fi
}%
\providecommand \@ifx [1]{%
 \ifx #1\expandafter \@firstoftwo
 \else \expandafter \@secondoftwo
 \fi
}%
\providecommand \natexlab [1]{#1}%
\providecommand \enquote  [1]{``#1''}%
\providecommand \bibnamefont  [1]{#1}%
\providecommand \bibfnamefont [1]{#1}%
\providecommand \citenamefont [1]{#1}%
\providecommand \href@noop [0]{\@secondoftwo}%
\providecommand \href [0]{\begingroup \@sanitize@url \@href}%
\providecommand \@href[1]{\@@startlink{#1}\@@href}%
\providecommand \@@href[1]{\endgroup#1\@@endlink}%
\providecommand \@sanitize@url [0]{\catcode `\\12\catcode `\$12\catcode
  `\&12\catcode `\#12\catcode `\^12\catcode `\_12\catcode `\%12\relax}%
\providecommand \@@startlink[1]{}%
\providecommand \@@endlink[0]{}%
\providecommand \url  [0]{\begingroup\@sanitize@url \@url }%
\providecommand \@url [1]{\endgroup\@href {#1}{\urlprefix }}%
\providecommand \urlprefix  [0]{URL }%
\providecommand \Eprint [0]{\href }%
\providecommand \doibase [0]{https://doi.org/}%
\providecommand \selectlanguage [0]{\@gobble}%
\providecommand \bibinfo  [0]{\@secondoftwo}%
\providecommand \bibfield  [0]{\@secondoftwo}%
\providecommand \translation [1]{[#1]}%
\providecommand \BibitemOpen [0]{}%
\providecommand \bibitemStop [0]{}%
\providecommand \bibitemNoStop [0]{.\EOS\space}%
\providecommand \EOS [0]{\spacefactor3000\relax}%
\providecommand \BibitemShut  [1]{\csname bibitem#1\endcsname}%
\let\auto@bib@innerbib\@empty
\bibitem [{\citenamefont {Vahala}(2003)}]{Vahala2003}%
  \BibitemOpen
  \bibfield  {author} {\bibinfo {author} {\bibfnamefont {K.~J.}\ \bibnamefont
  {Vahala}},\ }\bibfield  {title} {\bibinfo {title} {Optical microcavities},\
  }\href {https://doi.org/10.1038/nature01939} {\bibfield  {journal} {\bibinfo
  {journal} {Nature}\ }\textbf {\bibinfo {volume} {424}},\ \bibinfo {pages}
  {839} (\bibinfo {year} {2003})}\BibitemShut {NoStop}%
\bibitem [{\citenamefont {Tame}\ \emph {et~al.}(2013)\citenamefont {Tame},
  \citenamefont {McEnery}, \citenamefont {\"{O}zdemir}, \citenamefont {Lee},
  \citenamefont {Maier},\ and\ \citenamefont {Kim}}]{TameQuantumPlasmonics}%
  \BibitemOpen
  \bibfield  {author} {\bibinfo {author} {\bibfnamefont {M.~S.}\ \bibnamefont
  {Tame}}, \bibinfo {author} {\bibfnamefont {K.~R.}\ \bibnamefont {McEnery}},
  \bibinfo {author} {\bibfnamefont {{\c{S}}.~K.}\ \bibnamefont {\"{O}zdemir}},
  \bibinfo {author} {\bibfnamefont {J.}~\bibnamefont {Lee}}, \bibinfo {author}
  {\bibfnamefont {S.~A.}\ \bibnamefont {Maier}},\ and\ \bibinfo {author}
  {\bibfnamefont {M.~S.}\ \bibnamefont {Kim}},\ }\bibfield  {title} {\bibinfo
  {title} {Quantum plasmonics},\ }\href {https://doi.org/10.1038/nphys2615}
  {\bibfield  {journal} {\bibinfo  {journal} {Nature Physics}\ }\textbf
  {\bibinfo {volume} {9}},\ \bibinfo {pages} {329} (\bibinfo {year}
  {2013})}\BibitemShut {NoStop}%
\bibitem [{\citenamefont {Zhu}\ \emph {et~al.}(2016{\natexlab{a}})\citenamefont
  {Zhu}, \citenamefont {Esteban}, \citenamefont {Borisov}, \citenamefont
  {Baumberg}, \citenamefont {Nordlander}, \citenamefont {Lezec},\ and\
  \citenamefont {Aizpurua}}]{QuantumPlasmonicsReview}%
  \BibitemOpen
  \bibfield  {author} {\bibinfo {author} {\bibfnamefont {W.}~\bibnamefont
  {Zhu}}, \bibinfo {author} {\bibfnamefont {R.}~\bibnamefont {Esteban}},
  \bibinfo {author} {\bibfnamefont {A.~G.}\ \bibnamefont {Borisov}}, \bibinfo
  {author} {\bibfnamefont {J.~J.}\ \bibnamefont {Baumberg}}, \bibinfo {author}
  {\bibfnamefont {P.}~\bibnamefont {Nordlander}}, \bibinfo {author}
  {\bibfnamefont {H.~J.}\ \bibnamefont {Lezec}},\ and\ \bibinfo {author}
  {\bibfnamefont {K.~B.}\ \bibnamefont {Aizpurua}, \bibfnamefont {Javier
  an~Crozier}},\ }\bibfield  {title} {\bibinfo {title} {Quantum mechanical
  effects in plasmonic structures with subnanometre gaps},\ }\href
  {https://doi.org/10.1038/ncomms11495} {\bibfield  {journal} {\bibinfo
  {journal} {Nature Communications}\ }\textbf {\bibinfo {volume} {7}},\
  \bibinfo {pages} {11495} (\bibinfo {year} {2016}{\natexlab{a}})}\BibitemShut
  {NoStop}%
\bibitem [{\citenamefont {Chikkaraddy}\ \emph {et~al.}(2016)\citenamefont
  {Chikkaraddy}, \citenamefont {de~Nijs}, \citenamefont {Benz}, \citenamefont
  {Barrow}, \citenamefont {Scherman}, \citenamefont {Rosta}, \citenamefont
  {Demetriadou}, \citenamefont {Fox}, \citenamefont {Hess},\ and\ \citenamefont
  {Baumberg}}]{Chikkaraddy2016}%
  \BibitemOpen
  \bibfield  {author} {\bibinfo {author} {\bibfnamefont {R.}~\bibnamefont
  {Chikkaraddy}}, \bibinfo {author} {\bibfnamefont {B.}~\bibnamefont
  {de~Nijs}}, \bibinfo {author} {\bibfnamefont {F.}~\bibnamefont {Benz}},
  \bibinfo {author} {\bibfnamefont {S.~J.}\ \bibnamefont {Barrow}}, \bibinfo
  {author} {\bibfnamefont {O.~A.}\ \bibnamefont {Scherman}}, \bibinfo {author}
  {\bibfnamefont {E.}~\bibnamefont {Rosta}}, \bibinfo {author} {\bibfnamefont
  {A.}~\bibnamefont {Demetriadou}}, \bibinfo {author} {\bibfnamefont
  {P.}~\bibnamefont {Fox}}, \bibinfo {author} {\bibfnamefont {O.}~\bibnamefont
  {Hess}},\ and\ \bibinfo {author} {\bibfnamefont {J.~J.}\ \bibnamefont
  {Baumberg}},\ }\bibfield  {title} {\bibinfo {title} {Single-molecule strong
  coupling at room temperature in plasmonic nanocavities},\ }\href
  {https://doi.org/10.1038/nature17974} {\bibfield  {journal} {\bibinfo
  {journal} {Nature}\ }\textbf {\bibinfo {volume} {535}},\ \bibinfo {pages}
  {127} (\bibinfo {year} {2016})}\BibitemShut {NoStop}%
\bibitem [{\citenamefont {Xiao}\ \emph {et~al.}(2012)\citenamefont {Xiao},
  \citenamefont {Liu}, \citenamefont {Feng}, \citenamefont {Xu},\ and\
  \citenamefont {Yao}}]{xiao2012coupled}%
  \BibitemOpen
  \bibfield  {author} {\bibinfo {author} {\bibfnamefont {D.}~\bibnamefont
  {Xiao}}, \bibinfo {author} {\bibfnamefont {G.-B.}\ \bibnamefont {Liu}},
  \bibinfo {author} {\bibfnamefont {W.}~\bibnamefont {Feng}}, \bibinfo {author}
  {\bibfnamefont {X.}~\bibnamefont {Xu}},\ and\ \bibinfo {author}
  {\bibfnamefont {W.}~\bibnamefont {Yao}},\ }\bibfield  {title} {\bibinfo
  {title} {Coupled spin and valley physics in monolayers of mos 2 and other
  group-vi dichalcogenides},\ }{} {\bibfield  {journal} {\bibinfo
  {journal} {Physical Review Letters}\ }\textbf {\bibinfo {volume} {108}},\
  \bibinfo {pages} {196802} (\bibinfo {year} {2012})}\BibitemShut {NoStop}%
\bibitem [{\citenamefont {Korm{\'a}nyos}\ \emph {et~al.}(2015)\citenamefont
  {Korm{\'a}nyos}, \citenamefont {Burkard}, \citenamefont {Gmitra},
  \citenamefont {Fabian}, \citenamefont {Z{\'o}lyomi}, \citenamefont
  {Drummond},\ and\ \citenamefont {Fal’ko}}]{kormanyos2015k}%
  \BibitemOpen
  \bibfield  {author} {\bibinfo {author} {\bibfnamefont {A.}~\bibnamefont
  {Korm{\'a}nyos}}, \bibinfo {author} {\bibfnamefont {G.}~\bibnamefont
  {Burkard}}, \bibinfo {author} {\bibfnamefont {M.}~\bibnamefont {Gmitra}},
  \bibinfo {author} {\bibfnamefont {J.}~\bibnamefont {Fabian}}, \bibinfo
  {author} {\bibfnamefont {V.}~\bibnamefont {Z{\'o}lyomi}}, \bibinfo {author}
  {\bibfnamefont {N.~D.}\ \bibnamefont {Drummond}},\ and\ \bibinfo {author}
  {\bibfnamefont {V.}~\bibnamefont {Fal’ko}},\ }\bibfield  {title} {\bibinfo
  {title} {k{\textperiodcentered} p theory for two-dimensional transition metal
  dichalcogenide semiconductors},\ }{} {\bibfield  {journal}
  {\bibinfo  {journal} {2D Materials}\ }\textbf {\bibinfo {volume} {2}},\
  \bibinfo {pages} {022001} (\bibinfo {year} {2015})}\BibitemShut {NoStop}%
\bibitem [{\citenamefont {Chernikov}\ \emph {et~al.}(2014)\citenamefont
  {Chernikov}, \citenamefont {Berkelbach}, \citenamefont {Hill}, \citenamefont
  {Rigosi}, \citenamefont {Li}, \citenamefont {Aslan}, \citenamefont
  {Reichman}, \citenamefont {Hybertsen},\ and\ \citenamefont
  {Heinz}}]{chernikov2014exciton}%
  \BibitemOpen
  \bibfield  {author} {\bibinfo {author} {\bibfnamefont {A.}~\bibnamefont
  {Chernikov}}, \bibinfo {author} {\bibfnamefont {T.~C.}\ \bibnamefont
  {Berkelbach}}, \bibinfo {author} {\bibfnamefont {H.~M.}\ \bibnamefont
  {Hill}}, \bibinfo {author} {\bibfnamefont {A.}~\bibnamefont {Rigosi}},
  \bibinfo {author} {\bibfnamefont {Y.}~\bibnamefont {Li}}, \bibinfo {author}
  {\bibfnamefont {O.~B.}\ \bibnamefont {Aslan}}, \bibinfo {author}
  {\bibfnamefont {D.~R.}\ \bibnamefont {Reichman}}, \bibinfo {author}
  {\bibfnamefont {M.~S.}\ \bibnamefont {Hybertsen}},\ and\ \bibinfo {author}
  {\bibfnamefont {T.~F.}\ \bibnamefont {Heinz}},\ }\bibfield  {title} {\bibinfo
  {title} {Exciton binding energy and nonhydrogenic rydberg series in monolayer
  ws 2},\ }{} {\bibfield  {journal} {\bibinfo  {journal} {Physical
  Review Letters}\ }\textbf {\bibinfo {volume} {113}},\ \bibinfo {pages}
  {076802} (\bibinfo {year} {2014})}\BibitemShut {NoStop}%
\bibitem [{\citenamefont {Steinhoff}\ \emph {et~al.}(2017)\citenamefont
  {Steinhoff}, \citenamefont {Florian}, \citenamefont {R{\"o}sner},
  \citenamefont {Sch{\"o}nhoff}, \citenamefont {Wehling},\ and\ \citenamefont
  {Jahnke}}]{steinhoff2017exciton}%
  \BibitemOpen
  \bibfield  {author} {\bibinfo {author} {\bibfnamefont {A.}~\bibnamefont
  {Steinhoff}}, \bibinfo {author} {\bibfnamefont {M.}~\bibnamefont {Florian}},
  \bibinfo {author} {\bibfnamefont {M.}~\bibnamefont {R{\"o}sner}}, \bibinfo
  {author} {\bibfnamefont {G.}~\bibnamefont {Sch{\"o}nhoff}}, \bibinfo {author}
  {\bibfnamefont {T.}~\bibnamefont {Wehling}},\ and\ \bibinfo {author}
  {\bibfnamefont {F.}~\bibnamefont {Jahnke}},\ }\bibfield  {title} {\bibinfo
  {title} {Exciton fission in monolayer transition metal dichalcogenide
  semiconductors},\ }{} {\bibfield  {journal} {\bibinfo  {journal}
  {Nature Communications}\ }\textbf {\bibinfo {volume} {8}},\ \bibinfo {pages}
  {1166} (\bibinfo {year} {2017})}\BibitemShut {NoStop}%
\bibitem [{\citenamefont {Katsch}\ \emph {et~al.}(2020)\citenamefont {Katsch},
  \citenamefont {Selig},\ and\ \citenamefont {Knorr}}]{katsch2020exciton}%
  \BibitemOpen
  \bibfield  {author} {\bibinfo {author} {\bibfnamefont {F.}~\bibnamefont
  {Katsch}}, \bibinfo {author} {\bibfnamefont {M.}~\bibnamefont {Selig}},\ and\
  \bibinfo {author} {\bibfnamefont {A.}~\bibnamefont {Knorr}},\ }\bibfield
  {title} {\bibinfo {title} {Exciton-scattering-induced dephasing in
  two-dimensional semiconductors},\ }{} {\bibfield  {journal}
  {\bibinfo  {journal} {Physical Review Letters}\ }\textbf {\bibinfo {volume}
  {124}},\ \bibinfo {pages} {257402} (\bibinfo {year} {2020})}\BibitemShut
  {NoStop}%
\bibitem [{\citenamefont {Kern}\ \emph {et~al.}(2015)\citenamefont {Kern},
  \citenamefont {Tr\"{u}gler}, \citenamefont {Niehues}, \citenamefont
  {Ewering}, \citenamefont {Schmidt}, \citenamefont {Schneider}, \citenamefont
  {Najmaei}, \citenamefont {George}, \citenamefont {Zhang}, \citenamefont
  {Lou}, \citenamefont {Hohenester}, \citenamefont {Michaelis~de
  Vasconcellos},\ and\ \citenamefont {Bratschitsch}}]{kern-nanoantenna-2015}%
  \BibitemOpen
  \bibfield  {author} {\bibinfo {author} {\bibfnamefont {J.}~\bibnamefont
  {Kern}}, \bibinfo {author} {\bibfnamefont {A.}~\bibnamefont {Tr\"{u}gler}},
  \bibinfo {author} {\bibfnamefont {I.}~\bibnamefont {Niehues}}, \bibinfo
  {author} {\bibfnamefont {J.}~\bibnamefont {Ewering}}, \bibinfo {author}
  {\bibfnamefont {R.}~\bibnamefont {Schmidt}}, \bibinfo {author} {\bibfnamefont
  {R.}~\bibnamefont {Schneider}}, \bibinfo {author} {\bibfnamefont
  {S.}~\bibnamefont {Najmaei}}, \bibinfo {author} {\bibfnamefont
  {A.}~\bibnamefont {George}}, \bibinfo {author} {\bibfnamefont
  {J.}~\bibnamefont {Zhang}}, \bibinfo {author} {\bibfnamefont
  {J.}~\bibnamefont {Lou}}, \bibinfo {author} {\bibfnamefont {U.}~\bibnamefont
  {Hohenester}}, \bibinfo {author} {\bibfnamefont {S.}~\bibnamefont
  {Michaelis~de Vasconcellos}},\ and\ \bibinfo {author} {\bibfnamefont
  {R.}~\bibnamefont {Bratschitsch}},\ }\bibfield  {title} {\bibinfo {title}
  {Nanoantenna-enhanced light–matter interaction in atomically thin ws2},\
  }\href {https://doi.org/10.1021/acsphotonics.5b00123} {\bibfield  {journal}
  {\bibinfo  {journal} {ACS Photonics}\ }\textbf {\bibinfo {volume} {2}},\
  \bibinfo {pages} {1260} (\bibinfo {year} {2015})}\BibitemShut {NoStop}%
\bibitem [{\citenamefont {Zheng}\ \emph {et~al.}(2017)\citenamefont {Zheng},
  \citenamefont {Zhang}, \citenamefont {Deng}, \citenamefont {Kang},
  \citenamefont {Nordlander},\ and\ \citenamefont
  {Xu}}]{zheng-manipulating-2017}%
  \BibitemOpen
  \bibfield  {author} {\bibinfo {author} {\bibfnamefont {D.}~\bibnamefont
  {Zheng}}, \bibinfo {author} {\bibfnamefont {S.}~\bibnamefont {Zhang}},
  \bibinfo {author} {\bibfnamefont {Q.}~\bibnamefont {Deng}}, \bibinfo {author}
  {\bibfnamefont {M.}~\bibnamefont {Kang}}, \bibinfo {author} {\bibfnamefont
  {P.}~\bibnamefont {Nordlander}},\ and\ \bibinfo {author} {\bibfnamefont
  {H.}~\bibnamefont {Xu}},\ }\bibfield  {title} {\bibinfo {title} {Manipulating
  coherent plasmon–exciton interaction in a single silver nanorod on
  monolayer \uppercase{WS}e$_2$},\ }\href
  {https://doi.org/10.1021/acs.nanolett.7b01176} {\bibfield  {journal}
  {\bibinfo  {journal} {Nano Letters}\ }\textbf {\bibinfo {volume} {17}},\
  \bibinfo {pages} {3809} (\bibinfo {year} {2017})}\BibitemShut {NoStop}%
\bibitem [{\citenamefont {Wen}\ \emph {et~al.}(2017)\citenamefont {Wen},
  \citenamefont {Wang}, \citenamefont {Wang}, \citenamefont {Deng},
  \citenamefont {Zhuang}, \citenamefont {Zhang}, \citenamefont {Liu},
  \citenamefont {She}, \citenamefont {Chen}, \citenamefont {Chen},
  \citenamefont {Deng},\ and\ \citenamefont {Xu}}]{wen-room-temperature-2017}%
  \BibitemOpen
  \bibfield  {author} {\bibinfo {author} {\bibfnamefont {J.}~\bibnamefont
  {Wen}}, \bibinfo {author} {\bibfnamefont {H.}~\bibnamefont {Wang}}, \bibinfo
  {author} {\bibfnamefont {W.}~\bibnamefont {Wang}}, \bibinfo {author}
  {\bibfnamefont {Z.}~\bibnamefont {Deng}}, \bibinfo {author} {\bibfnamefont
  {C.}~\bibnamefont {Zhuang}}, \bibinfo {author} {\bibfnamefont
  {Y.}~\bibnamefont {Zhang}}, \bibinfo {author} {\bibfnamefont
  {F.}~\bibnamefont {Liu}}, \bibinfo {author} {\bibfnamefont {J.}~\bibnamefont
  {She}}, \bibinfo {author} {\bibfnamefont {J.}~\bibnamefont {Chen}}, \bibinfo
  {author} {\bibfnamefont {H.}~\bibnamefont {Chen}}, \bibinfo {author}
  {\bibfnamefont {S.}~\bibnamefont {Deng}},\ and\ \bibinfo {author}
  {\bibfnamefont {N.}~\bibnamefont {Xu}},\ }\bibfield  {title} {\bibinfo
  {title} {Room-temperature strong light–matter interaction with active
  control in single plasmonic nanorod coupled with two-dimensional atomic
  crystals},\ }\href {https://doi.org/10.1021/acs.nanolett.7b01344} {\bibfield
  {journal} {\bibinfo  {journal} {Nano Letters}\ }\textbf {\bibinfo {volume}
  {17}},\ \bibinfo {pages} {4689} (\bibinfo {year} {2017})}\BibitemShut
  {NoStop}%
\bibitem [{\citenamefont {St\"{u}hrenberg}\ \emph {et~al.}(2018)\citenamefont
  {St\"{u}hrenberg}, \citenamefont {Munkhbat}, \citenamefont {Baranov},
  \citenamefont {Cuadra}, \citenamefont {Yankovich}, \citenamefont
  {Antosiewicz}, \citenamefont {Olsson},\ and\ \citenamefont
  {Shegai}}]{struhrenberg-strong-2018}%
  \BibitemOpen
  \bibfield  {author} {\bibinfo {author} {\bibfnamefont {M.}~\bibnamefont
  {St\"{u}hrenberg}}, \bibinfo {author} {\bibfnamefont {B.}~\bibnamefont
  {Munkhbat}}, \bibinfo {author} {\bibfnamefont {D.~G.}\ \bibnamefont
  {Baranov}}, \bibinfo {author} {\bibfnamefont {J.}~\bibnamefont {Cuadra}},
  \bibinfo {author} {\bibfnamefont {A.~B.}\ \bibnamefont {Yankovich}}, \bibinfo
  {author} {\bibfnamefont {T.~J.}\ \bibnamefont {Antosiewicz}}, \bibinfo
  {author} {\bibfnamefont {E.}~\bibnamefont {Olsson}},\ and\ \bibinfo {author}
  {\bibfnamefont {T.}~\bibnamefont {Shegai}},\ }\bibfield  {title} {\bibinfo
  {title} {Strong light–matter coupling between plasmons in individual gold
  bi-pyramids and excitons in mono- and multilayer \uppercase{WS}e$_2$},\
  }\href {https://doi.org/10.1021/acs.nanolett.8b02652} {\bibfield  {journal}
  {\bibinfo  {journal} {Nano Letters}\ }\textbf {\bibinfo {volume} {18}},\
  \bibinfo {pages} {5938} (\bibinfo {year} {2018})}\BibitemShut {NoStop}%
\bibitem [{\citenamefont {Geisler}\ \emph {et~al.}(2019)\citenamefont
  {Geisler}, \citenamefont {Cui}, \citenamefont {Wang}, \citenamefont
  {Rindzevicius}, \citenamefont {Gammelgaard}, \citenamefont {Jessen},
  \citenamefont {Gon\c{c}alves}, \citenamefont {Todisco}, \citenamefont
  {B\o{o}ggild}, \citenamefont {Boisen}, \citenamefont {Wubs}, \citenamefont
  {Mortensen}, \citenamefont {Xiao},\ and\ \citenamefont
  {Stenger}}]{geisler-single-2019}%
  \BibitemOpen
  \bibfield  {author} {\bibinfo {author} {\bibfnamefont {M.}~\bibnamefont
  {Geisler}}, \bibinfo {author} {\bibfnamefont {X.}~\bibnamefont {Cui}},
  \bibinfo {author} {\bibfnamefont {J.}~\bibnamefont {Wang}}, \bibinfo {author}
  {\bibfnamefont {T.}~\bibnamefont {Rindzevicius}}, \bibinfo {author}
  {\bibfnamefont {L.}~\bibnamefont {Gammelgaard}}, \bibinfo {author}
  {\bibfnamefont {B.~S.}\ \bibnamefont {Jessen}}, \bibinfo {author}
  {\bibfnamefont {P.~A.~D.}\ \bibnamefont {Gon\c{c}alves}}, \bibinfo {author}
  {\bibfnamefont {F.}~\bibnamefont {Todisco}}, \bibinfo {author} {\bibfnamefont
  {P.}~\bibnamefont {B\o{o}ggild}}, \bibinfo {author} {\bibfnamefont
  {A.}~\bibnamefont {Boisen}}, \bibinfo {author} {\bibfnamefont
  {M.}~\bibnamefont {Wubs}}, \bibinfo {author} {\bibfnamefont {N.~A.}\
  \bibnamefont {Mortensen}}, \bibinfo {author} {\bibfnamefont {S.}~\bibnamefont
  {Xiao}},\ and\ \bibinfo {author} {\bibfnamefont {N.}~\bibnamefont
  {Stenger}},\ }\bibfield  {title} {\bibinfo {title} {Single-crystalline gold
  nanodisks on \uppercase{WS}$_2$ mono- and multilayers for strong coupling at
  room temperature},\ }\href {https://doi.org/10.1021/acsphotonics.8b01766}
  {\bibfield  {journal} {\bibinfo  {journal} {ACS Photonics}\ }\textbf
  {\bibinfo {volume} {6}},\ \bibinfo {pages} {994} (\bibinfo {year}
  {2019})}\BibitemShut {NoStop}%
\bibitem [{\citenamefont {Kleemann}\ \emph {et~al.}(2017)\citenamefont
  {Kleemann}, \citenamefont {Chikkaraddy}, \citenamefont {Alexeev},
  \citenamefont {Kos}, \citenamefont {Carnegie}, \citenamefont {Deacon},
  \citenamefont {de~Pury}, \citenamefont {Gro$\beta$e}, \citenamefont
  {de~Nijs}, \citenamefont {Mertens}, \citenamefont {Tartakovskii},\ and\
  \citenamefont {Baumberg}}]{kleeman-strong-2017}%
  \BibitemOpen
  \bibfield  {author} {\bibinfo {author} {\bibfnamefont {M.}~\bibnamefont
  {Kleemann}}, \bibinfo {author} {\bibfnamefont {R.}~\bibnamefont
  {Chikkaraddy}}, \bibinfo {author} {\bibfnamefont {E.}~\bibnamefont
  {Alexeev}}, \bibinfo {author} {\bibfnamefont {D.}~\bibnamefont {Kos}},
  \bibinfo {author} {\bibfnamefont {C.}~\bibnamefont {Carnegie}}, \bibinfo
  {author} {\bibfnamefont {W.}~\bibnamefont {Deacon}}, \bibinfo {author}
  {\bibfnamefont {A.}~\bibnamefont {de~Pury}}, \bibinfo {author} {\bibfnamefont
  {C.}~\bibnamefont {Gro$\beta$e}}, \bibinfo {author} {\bibfnamefont
  {B.}~\bibnamefont {de~Nijs}}, \bibinfo {author} {\bibfnamefont
  {J.}~\bibnamefont {Mertens}}, \bibinfo {author} {\bibfnamefont
  {A.}~\bibnamefont {Tartakovskii}},\ and\ \bibinfo {author} {\bibfnamefont
  {J.}~\bibnamefont {Baumberg}},\ }\bibfield  {title} {\bibinfo {title}
  {Strong-coupling of \uppercase{WS}e$_2$ in ultra-compact plasmonic
  nanocavities at room temperature},\ }\href
  {https://doi.org/https://doi.org/10.1038/s41467-017-01398-3} {\bibfield
  {journal} {\bibinfo  {journal} {Nat Commun}\ }\textbf {\bibinfo {volume}
  {8}},\ \bibinfo {pages} {1296} (\bibinfo {year} {2017})}\BibitemShut
  {NoStop}%
\bibitem [{\citenamefont {Liu}\ \emph {et~al.}(2020)\citenamefont {Liu},
  \citenamefont {Tobing}, \citenamefont {Yu}, \citenamefont {Tong},
  \citenamefont {Qiang}, \citenamefont {Fern\'andez-Dom\'inguez}, \citenamefont
  {Garcia-Vidal}, \citenamefont {Zhang}, \citenamefont {Wang},\ and\
  \citenamefont {Luo}}]{liu-strong-2020}%
  \BibitemOpen
  \bibfield  {author} {\bibinfo {author} {\bibfnamefont {L.}~\bibnamefont
  {Liu}}, \bibinfo {author} {\bibfnamefont {L.~Y.~M.}\ \bibnamefont {Tobing}},
  \bibinfo {author} {\bibfnamefont {X.}~\bibnamefont {Yu}}, \bibinfo {author}
  {\bibfnamefont {J.}~\bibnamefont {Tong}}, \bibinfo {author} {\bibfnamefont
  {B.}~\bibnamefont {Qiang}}, \bibinfo {author} {\bibfnamefont {A.~I.}\
  \bibnamefont {Fern\'andez-Dom\'inguez}}, \bibinfo {author} {\bibfnamefont
  {F.~J.}\ \bibnamefont {Garcia-Vidal}}, \bibinfo {author} {\bibfnamefont
  {D.~H.}\ \bibnamefont {Zhang}}, \bibinfo {author} {\bibfnamefont {Q.~J.}\
  \bibnamefont {Wang}},\ and\ \bibinfo {author} {\bibfnamefont
  {Y.}~\bibnamefont {Luo}},\ }\bibfield  {title} {\bibinfo {title} {Strong
  plasmon–exciton interactions on nanoantenna array–monolayer ws2 hybrid
  system},\ }\href {https://doi.org/https://doi.org/10.1002/adom.201901002}
  {\bibfield  {journal} {\bibinfo  {journal} {Advanced Optical Materials}\
  }\textbf {\bibinfo {volume} {8}},\ \bibinfo {pages} {1901002} (\bibinfo
  {year} {2020})}\BibitemShut {NoStop}%
\bibitem [{\citenamefont {Abid}\ \emph {et~al.}(2017)\citenamefont {Abid},
  \citenamefont {Chen}, \citenamefont {Yuan}, \citenamefont {Bohloul},
  \citenamefont {Najmaei}, \citenamefont {Avendano}, \citenamefont {P\'echou},
  \citenamefont {Mlayah},\ and\ \citenamefont {Lou}}]{abid-temperature-2017}%
  \BibitemOpen
  \bibfield  {author} {\bibinfo {author} {\bibfnamefont {I.}~\bibnamefont
  {Abid}}, \bibinfo {author} {\bibfnamefont {W.}~\bibnamefont {Chen}}, \bibinfo
  {author} {\bibfnamefont {J.}~\bibnamefont {Yuan}}, \bibinfo {author}
  {\bibfnamefont {A.}~\bibnamefont {Bohloul}}, \bibinfo {author} {\bibfnamefont
  {S.}~\bibnamefont {Najmaei}}, \bibinfo {author} {\bibfnamefont
  {C.}~\bibnamefont {Avendano}}, \bibinfo {author} {\bibfnamefont
  {R.}~\bibnamefont {P\'echou}}, \bibinfo {author} {\bibfnamefont
  {A.}~\bibnamefont {Mlayah}},\ and\ \bibinfo {author} {\bibfnamefont
  {J.}~\bibnamefont {Lou}},\ }\bibfield  {title} {\bibinfo {title}
  {Temperature-dependent plasmon–exciton interactions in hybrid
  \uppercase{A}u/\uppercase{M}o\uppercase{S}e$_2$ nanostructures},\ }\href
  {https://doi.org/10.1021/acsphotonics.6b00957} {\bibfield  {journal}
  {\bibinfo  {journal} {ACS Photonics}\ }\textbf {\bibinfo {volume} {4}},\
  \bibinfo {pages} {1653} (\bibinfo {year} {2017})}\BibitemShut {NoStop}%
\bibitem [{\citenamefont {Abid}\ \emph {et~al.}(2018)\citenamefont {Abid},
  \citenamefont {Chen}, \citenamefont {Yuan}, \citenamefont {Najmaei},
  \citenamefont {Pe{\~n}afiel}, \citenamefont {P{\'e}chou}, \citenamefont
  {Large}, \citenamefont {Lou},\ and\ \citenamefont
  {Mlayah}}]{abid-surface-2018}%
  \BibitemOpen
  \bibfield  {author} {\bibinfo {author} {\bibfnamefont {I.}~\bibnamefont
  {Abid}}, \bibinfo {author} {\bibfnamefont {W.}~\bibnamefont {Chen}}, \bibinfo
  {author} {\bibfnamefont {J.}~\bibnamefont {Yuan}}, \bibinfo {author}
  {\bibfnamefont {S.}~\bibnamefont {Najmaei}}, \bibinfo {author} {\bibfnamefont
  {E.~C.}\ \bibnamefont {Pe{\~n}afiel}}, \bibinfo {author} {\bibfnamefont
  {R.}~\bibnamefont {P{\'e}chou}}, \bibinfo {author} {\bibfnamefont
  {N.}~\bibnamefont {Large}}, \bibinfo {author} {\bibfnamefont
  {J.}~\bibnamefont {Lou}},\ and\ \bibinfo {author} {\bibfnamefont
  {A.}~\bibnamefont {Mlayah}},\ }\bibfield  {title} {\bibinfo {title} {Surface
  enhanced resonant raman scattering in hybrid mose$_2$@au nanostructures},\
  }\href {https://doi.org/10.1364/OE.26.029411} {\bibfield  {journal} {\bibinfo
   {journal} {Opt. Express}\ }\textbf {\bibinfo {volume} {26}},\ \bibinfo
  {pages} {29411} (\bibinfo {year} {2018})}\BibitemShut {NoStop}%
\bibitem [{\citenamefont {Bisht}\ \emph {et~al.}(2019)\citenamefont {Bisht},
  \citenamefont {Cuadra}, \citenamefont {Wers\"all}, \citenamefont {Canales},
  \citenamefont {Antosiewicz},\ and\ \citenamefont
  {Shegai}}]{bisht-collective-2019}%
  \BibitemOpen
  \bibfield  {author} {\bibinfo {author} {\bibfnamefont {A.}~\bibnamefont
  {Bisht}}, \bibinfo {author} {\bibfnamefont {J.}~\bibnamefont {Cuadra}},
  \bibinfo {author} {\bibfnamefont {M.}~\bibnamefont {Wers\"all}}, \bibinfo
  {author} {\bibfnamefont {A.}~\bibnamefont {Canales}}, \bibinfo {author}
  {\bibfnamefont {T.~J.}\ \bibnamefont {Antosiewicz}},\ and\ \bibinfo {author}
  {\bibfnamefont {T.}~\bibnamefont {Shegai}},\ }\bibfield  {title} {\bibinfo
  {title} {Collective strong light-matter coupling in hierarchical
  microcavity-plasmon-exciton systems},\ }\href
  {https://doi.org/10.1021/acs.nanolett.8b03639} {\bibfield  {journal}
  {\bibinfo  {journal} {Nano Letters}\ }\textbf {\bibinfo {volume} {19}},\
  \bibinfo {pages} {189} (\bibinfo {year} {2019})}\BibitemShut {NoStop}%
\bibitem [{\citenamefont {Schneider}\ \emph {et~al.}(2018)\citenamefont
  {Schneider}, \citenamefont {Glazov}, \citenamefont {Korn}, \citenamefont
  {H\"{o}fling},\ and\ \citenamefont
  {Urbaszek}}]{schneider-two-dimensional-2018}%
  \BibitemOpen
  \bibfield  {author} {\bibinfo {author} {\bibfnamefont {C.}~\bibnamefont
  {Schneider}}, \bibinfo {author} {\bibfnamefont {M.}~\bibnamefont {Glazov}},
  \bibinfo {author} {\bibfnamefont {T.}~\bibnamefont {Korn}}, \bibinfo {author}
  {\bibfnamefont {S.}~\bibnamefont {H\"{o}fling}},\ and\ \bibinfo {author}
  {\bibfnamefont {B.}~\bibnamefont {Urbaszek}},\ }\bibfield  {title} {\bibinfo
  {title} {Two-dimensional semiconductors in the regime of strong light-matter
  coupling},\ }\href
  {https://doi.org/https://doi.org/10.1038/s41467-018-04866-6} {\bibfield
  {journal} {\bibinfo  {journal} {Nature Communincations}\ }\textbf {\bibinfo
  {volume} {9}},\ \bibinfo {pages} {2695} (\bibinfo {year} {2018})}\BibitemShut
  {NoStop}%
\bibitem [{\citenamefont {Weisbuch}\ \emph {et~al.}(1992)\citenamefont
  {Weisbuch}, \citenamefont {Nishioka}, \citenamefont {Ishikawa},\ and\
  \citenamefont {Arakawa}}]{PhysRevLett.69.3314}%
  \BibitemOpen
  \bibfield  {author} {\bibinfo {author} {\bibfnamefont {C.}~\bibnamefont
  {Weisbuch}}, \bibinfo {author} {\bibfnamefont {M.}~\bibnamefont {Nishioka}},
  \bibinfo {author} {\bibfnamefont {A.}~\bibnamefont {Ishikawa}},\ and\
  \bibinfo {author} {\bibfnamefont {Y.}~\bibnamefont {Arakawa}},\ }\bibfield
  {title} {\bibinfo {title} {Observation of the coupled exciton-photon mode
  splitting in a semiconductor quantum microcavity},\ }\href
  {https://doi.org/10.1103/PhysRevLett.69.3314} {\bibfield  {journal} {\bibinfo
   {journal} {Phys. Rev. Lett.}\ }\textbf {\bibinfo {volume} {69}},\ \bibinfo
  {pages} {3314} (\bibinfo {year} {1992})}\BibitemShut {NoStop}%
\bibitem [{\citenamefont {Reithmaier}\ \emph {et~al.}(2004)\citenamefont
  {Reithmaier}, \citenamefont {S\c{e}k}, \citenamefont {L\"{o}ffler},
  \citenamefont {Hofmann}, \citenamefont {Kuhn}, \citenamefont {Reitzenstein},
  \citenamefont {Keldysh}, \citenamefont {Kulakovskii}, \citenamefont
  {Reinecke},\ and\ \citenamefont {Forchel}}]{reithmaier-strong-2004}%
  \BibitemOpen
  \bibfield  {author} {\bibinfo {author} {\bibfnamefont {J.}~\bibnamefont
  {Reithmaier}}, \bibinfo {author} {\bibfnamefont {G.}~\bibnamefont {S\c{e}k}},
  \bibinfo {author} {\bibfnamefont {A.}~\bibnamefont {L\"{o}ffler}}, \bibinfo
  {author} {\bibfnamefont {C.}~\bibnamefont {Hofmann}}, \bibinfo {author}
  {\bibfnamefont {S.}~\bibnamefont {Kuhn}}, \bibinfo {author} {\bibfnamefont
  {S.}~\bibnamefont {Reitzenstein}}, \bibinfo {author} {\bibfnamefont {L.~V.}\
  \bibnamefont {Keldysh}}, \bibinfo {author} {\bibfnamefont {V.~D.}\
  \bibnamefont {Kulakovskii}}, \bibinfo {author} {\bibfnamefont {T.~L.}\
  \bibnamefont {Reinecke}},\ and\ \bibinfo {author} {\bibfnamefont
  {A.}~\bibnamefont {Forchel}},\ }\bibfield  {title} {\bibinfo {title} {Strong
  coupling in a single quantum dot–semiconductor microcavity system},\ }\href
  {https://doi.org/https://doi.org/10.1038/nature02969} {\bibfield  {journal}
  {\bibinfo  {journal} {Nature}\ }\textbf {\bibinfo {volume} {432}},\ \bibinfo
  {pages} {197–200} (\bibinfo {year} {2004})}\BibitemShut {NoStop}%
\bibitem [{\citenamefont {Yoshie}\ \emph {et~al.}(2004)\citenamefont {Yoshie},
  \citenamefont {Scherer}, \citenamefont {Hendrickson}, \citenamefont
  {Khitrova}, \citenamefont {Gibbs}, \citenamefont {Rupper}, \citenamefont
  {Ell}, \citenamefont {Shchekin},\ and\ \citenamefont {Deppe}}]{Yoshie2004}%
  \BibitemOpen
  \bibfield  {author} {\bibinfo {author} {\bibfnamefont {T.}~\bibnamefont
  {Yoshie}}, \bibinfo {author} {\bibfnamefont {A.}~\bibnamefont {Scherer}},
  \bibinfo {author} {\bibfnamefont {J.}~\bibnamefont {Hendrickson}}, \bibinfo
  {author} {\bibfnamefont {G.}~\bibnamefont {Khitrova}}, \bibinfo {author}
  {\bibfnamefont {H.~M.}\ \bibnamefont {Gibbs}}, \bibinfo {author}
  {\bibfnamefont {G.}~\bibnamefont {Rupper}}, \bibinfo {author} {\bibfnamefont
  {C.}~\bibnamefont {Ell}}, \bibinfo {author} {\bibfnamefont {O.~B.}\
  \bibnamefont {Shchekin}},\ and\ \bibinfo {author} {\bibfnamefont {D.~G.}\
  \bibnamefont {Deppe}},\ }\bibfield  {title} {\bibinfo {title} {Vacuum rabi
  splitting with a single quantum dot in a photonic crystal nanocavity},\
  }\href {https://doi.org/10.1038/nature03119} {\bibfield  {journal} {\bibinfo
  {journal} {Nature}\ }\textbf {\bibinfo {volume} {432}},\ \bibinfo {pages}
  {200} (\bibinfo {year} {2004})}\BibitemShut {NoStop}%
\bibitem [{\citenamefont {Peter}\ \emph {et~al.}(2005)\citenamefont {Peter},
  \citenamefont {Senellart}, \citenamefont {Martrou}, \citenamefont
  {Lema\^{\i}tre}, \citenamefont {Hours}, \citenamefont {G\'erard},\ and\
  \citenamefont {Bloch}}]{peter-exciton-2005}%
  \BibitemOpen
  \bibfield  {author} {\bibinfo {author} {\bibfnamefont {E.}~\bibnamefont
  {Peter}}, \bibinfo {author} {\bibfnamefont {P.}~\bibnamefont {Senellart}},
  \bibinfo {author} {\bibfnamefont {D.}~\bibnamefont {Martrou}}, \bibinfo
  {author} {\bibfnamefont {A.}~\bibnamefont {Lema\^{\i}tre}}, \bibinfo {author}
  {\bibfnamefont {J.}~\bibnamefont {Hours}}, \bibinfo {author} {\bibfnamefont
  {J.~M.}\ \bibnamefont {G\'erard}},\ and\ \bibinfo {author} {\bibfnamefont
  {J.}~\bibnamefont {Bloch}},\ }\bibfield  {title} {\bibinfo {title}
  {Exciton-photon strong-coupling regime for a single quantum dot embedded in a
  microcavity},\ }\href {https://doi.org/10.1103/PhysRevLett.95.067401}
  {\bibfield  {journal} {\bibinfo  {journal} {Phys. Rev. Lett.}\ }\textbf
  {\bibinfo {volume} {95}},\ \bibinfo {pages} {067401} (\bibinfo {year}
  {2005})}\BibitemShut {NoStop}%
\bibitem [{\citenamefont {Kristensen}\ and\ \citenamefont
  {Hughes}(2014)}]{kristensen-modes-2014}%
  \BibitemOpen
  \bibfield  {author} {\bibinfo {author} {\bibfnamefont {P.~T.}\ \bibnamefont
  {Kristensen}}\ and\ \bibinfo {author} {\bibfnamefont {S.}~\bibnamefont
  {Hughes}},\ }\bibfield  {title} {{\selectlanguage {en}\bibinfo {title} {Modes
  and {Mode} {Volumes} of {Leaky} {Optical} {Cavities} and {Plasmonic}
  {Nanoresonators}}},\ }\href {https://doi.org/10.1021/ph400114e} {\bibfield
  {journal} {\bibinfo  {journal} {ACS Photonics}\ }\textbf {\bibinfo {volume}
  {1}},\ \bibinfo {pages} {2} (\bibinfo {year} {2014})}\BibitemShut {NoStop}%
\bibitem [{\citenamefont {Zschiedrich}\ \emph {et~al.}(2018)\citenamefont
  {Zschiedrich}, \citenamefont {Binkowski}, \citenamefont {Nikolay},
  \citenamefont {Benson}, \citenamefont {Kewes},\ and\ \citenamefont
  {Burger}}]{PhysRevA.98.043806}%
  \BibitemOpen
  \bibfield  {author} {\bibinfo {author} {\bibfnamefont {L.}~\bibnamefont
  {Zschiedrich}}, \bibinfo {author} {\bibfnamefont {F.}~\bibnamefont
  {Binkowski}}, \bibinfo {author} {\bibfnamefont {N.}~\bibnamefont {Nikolay}},
  \bibinfo {author} {\bibfnamefont {O.}~\bibnamefont {Benson}}, \bibinfo
  {author} {\bibfnamefont {G.}~\bibnamefont {Kewes}},\ and\ \bibinfo {author}
  {\bibfnamefont {S.}~\bibnamefont {Burger}},\ }\bibfield  {title} {\bibinfo
  {title} {Riesz-projection-based theory of light-matter interaction in
  dispersive nanoresonators},\ }\href
  {https://doi.org/10.1103/PhysRevA.98.043806} {\bibfield  {journal} {\bibinfo
  {journal} {Phys. Rev. A}\ }\textbf {\bibinfo {volume} {98}},\ \bibinfo
  {pages} {043806} (\bibinfo {year} {2018})}\BibitemShut {NoStop}%
\bibitem [{\citenamefont {Muljarov}\ and\ \citenamefont
  {Langbein}(2016)}]{PhysRevB.94.235438}%
  \BibitemOpen
  \bibfield  {author} {\bibinfo {author} {\bibfnamefont {E.~A.}\ \bibnamefont
  {Muljarov}}\ and\ \bibinfo {author} {\bibfnamefont {W.}~\bibnamefont
  {Langbein}},\ }\bibfield  {title} {\bibinfo {title} {Exact mode volume and
  purcell factor of open optical systems},\ }\href
  {https://doi.org/10.1103/PhysRevB.94.235438} {\bibfield  {journal} {\bibinfo
  {journal} {Phys. Rev. B}\ }\textbf {\bibinfo {volume} {94}},\ \bibinfo
  {pages} {235438} (\bibinfo {year} {2016})}\BibitemShut {NoStop}%
\bibitem [{\citenamefont {Lalanne}\ \emph {et~al.}(2018)\citenamefont
  {Lalanne}, \citenamefont {Yan}, \citenamefont {Vynck}, \citenamefont
  {Sauvan},\ and\ \citenamefont {Hugonin}}]{lalanne-light-2018}%
  \BibitemOpen
  \bibfield  {author} {\bibinfo {author} {\bibfnamefont {P.}~\bibnamefont
  {Lalanne}}, \bibinfo {author} {\bibfnamefont {W.}~\bibnamefont {Yan}},
  \bibinfo {author} {\bibfnamefont {K.}~\bibnamefont {Vynck}}, \bibinfo
  {author} {\bibfnamefont {C.}~\bibnamefont {Sauvan}},\ and\ \bibinfo {author}
  {\bibfnamefont {J.-P.}\ \bibnamefont {Hugonin}},\ }\bibfield  {title}
  {\bibinfo {title} {Light interaction with photonic and plasmonic
  resonances},\ }\href {https://doi.org/10.1002/lpor.201700113} {\bibfield
  {journal} {\bibinfo  {journal} {Laser \& Photonics Reviews}\ }\textbf
  {\bibinfo {volume} {12}},\ \bibinfo {pages} {1700113} (\bibinfo {year}
  {2018})}\BibitemShut {NoStop}%
\bibitem [{\citenamefont {Kristensen}\ \emph {et~al.}(2020)\citenamefont
  {Kristensen}, \citenamefont {Herrmann}, \citenamefont {Intravaia},\ and\
  \citenamefont {Busch}}]{kristensen-modeling-2020}%
  \BibitemOpen
  \bibfield  {author} {\bibinfo {author} {\bibfnamefont {P.~T.}\ \bibnamefont
  {Kristensen}}, \bibinfo {author} {\bibfnamefont {K.}~\bibnamefont
  {Herrmann}}, \bibinfo {author} {\bibfnamefont {F.}~\bibnamefont
  {Intravaia}},\ and\ \bibinfo {author} {\bibfnamefont {K.}~\bibnamefont
  {Busch}},\ }\bibfield  {title} {\bibinfo {title} {Modeling electromagnetic
  resonators using quasinormal modes},\ }\href
  {https://doi.org/10.1364/AOP.377940} {\bibfield  {journal} {\bibinfo
  {journal} {Advances in Optics and Photonics}\ }\textbf {\bibinfo {volume}
  {12}},\ \bibinfo {pages} {612} (\bibinfo {year} {2020})}\BibitemShut
  {NoStop}%
\bibitem [{\citenamefont {Selig}\ \emph {et~al.}(2016)\citenamefont {Selig},
  \citenamefont {Bergh{\"a}user}, \citenamefont {Raja}, \citenamefont {Nagler},
  \citenamefont {Sch{\"u}ller}, \citenamefont {Heinz}, \citenamefont {Korn},
  \citenamefont {Chernikov}, \citenamefont {Mali\'{c}},\ and\ \citenamefont
  {Knorr}}]{selig2016excitonic}%
  \BibitemOpen
  \bibfield  {author} {\bibinfo {author} {\bibfnamefont {M.}~\bibnamefont
  {Selig}}, \bibinfo {author} {\bibfnamefont {G.}~\bibnamefont
  {Bergh{\"a}user}}, \bibinfo {author} {\bibfnamefont {A.}~\bibnamefont
  {Raja}}, \bibinfo {author} {\bibfnamefont {P.}~\bibnamefont {Nagler}},
  \bibinfo {author} {\bibfnamefont {C.}~\bibnamefont {Sch{\"u}ller}}, \bibinfo
  {author} {\bibfnamefont {T.~F.}\ \bibnamefont {Heinz}}, \bibinfo {author}
  {\bibfnamefont {T.}~\bibnamefont {Korn}}, \bibinfo {author} {\bibfnamefont
  {A.}~\bibnamefont {Chernikov}}, \bibinfo {author} {\bibfnamefont
  {E.}~\bibnamefont {Mali\'{c}}},\ and\ \bibinfo {author} {\bibfnamefont
  {A.}~\bibnamefont {Knorr}},\ }\bibfield  {title} {\bibinfo {title} {Excitonic
  linewidth and coherence lifetime in monolayer transition metal
  dichalcogenides},\ } {} {\bibfield  {journal} {\bibinfo  {journal}
  {Nature Communications}\ }\textbf {\bibinfo {volume} {7}},\ \bibinfo {pages}
  {13279} (\bibinfo {year} {2016})}\BibitemShut {NoStop}%
\bibitem [{\citenamefont {Khatibi}\ \emph {et~al.}(2018)\citenamefont
  {Khatibi}, \citenamefont {Feierabend}, \citenamefont {Selig}, \citenamefont
  {Brem}, \citenamefont {Linder{\"a}lv}, \citenamefont {Erhart},\ and\
  \citenamefont {Mali\'{c}}}]{khatibi2018impact}%
  \BibitemOpen
  \bibfield  {author} {\bibinfo {author} {\bibfnamefont {Z.}~\bibnamefont
  {Khatibi}}, \bibinfo {author} {\bibfnamefont {M.}~\bibnamefont {Feierabend}},
  \bibinfo {author} {\bibfnamefont {M.}~\bibnamefont {Selig}}, \bibinfo
  {author} {\bibfnamefont {S.}~\bibnamefont {Brem}}, \bibinfo {author}
  {\bibfnamefont {C.}~\bibnamefont {Linder{\"a}lv}}, \bibinfo {author}
  {\bibfnamefont {P.}~\bibnamefont {Erhart}},\ and\ \bibinfo {author}
  {\bibfnamefont {E.}~\bibnamefont {Mali\'{c}}},\ }\bibfield  {title} {\bibinfo
  {title} {Impact of strain on the excitonic linewidth in transition metal
  dichalcogenides},\ }{} {\bibfield  {journal} {\bibinfo  {journal}
  {2D Materials}\ }\textbf {\bibinfo {volume} {6}},\ \bibinfo {pages} {015015}
  (\bibinfo {year} {2018})}\BibitemShut {NoStop}%
\bibitem [{\citenamefont {Zhu}\ \emph {et~al.}(1990)\citenamefont {Zhu},
  \citenamefont {Gauthier}, \citenamefont {Morin}, \citenamefont {Wu},
  \citenamefont {Carmichael},\ and\ \citenamefont
  {Mossberg}}]{PhysRevLett.64.2499}%
  \BibitemOpen
  \bibfield  {author} {\bibinfo {author} {\bibfnamefont {Y.}~\bibnamefont
  {Zhu}}, \bibinfo {author} {\bibfnamefont {D.~J.}\ \bibnamefont {Gauthier}},
  \bibinfo {author} {\bibfnamefont {S.~E.}\ \bibnamefont {Morin}}, \bibinfo
  {author} {\bibfnamefont {Q.}~\bibnamefont {Wu}}, \bibinfo {author}
  {\bibfnamefont {H.~J.}\ \bibnamefont {Carmichael}},\ and\ \bibinfo {author}
  {\bibfnamefont {T.~W.}\ \bibnamefont {Mossberg}},\ }\bibfield  {title}
  {\bibinfo {title} {Vacuum rabi splitting as a feature of linear-dispersion
  theory: Analysis and experimental observations},\ }\href
  {https://doi.org/10.1103/PhysRevLett.64.2499} {\bibfield  {journal} {\bibinfo
   {journal} {Phys. Rev. Lett.}\ }\textbf {\bibinfo {volume} {64}},\ \bibinfo
  {pages} {2499} (\bibinfo {year} {1990})}\BibitemShut {NoStop}%
\bibitem [{\citenamefont {Hu}\ and\ \citenamefont
  {Fei}(2020)}]{hu-recent-2020}%
  \BibitemOpen
  \bibfield  {author} {\bibinfo {author} {\bibfnamefont {F.}~\bibnamefont
  {Hu}}\ and\ \bibinfo {author} {\bibfnamefont {Z.}~\bibnamefont {Fei}},\
  }\bibfield  {title} {\bibinfo {title} {Recent progress on exciton polaritons
  in layered transition-metal dichalcogenides},\ }\href
  {https://doi.org/10.1002/adom.201901003} {\bibfield  {journal} {\bibinfo
  {journal} {Advanced Optical Materials}\ }\textbf {\bibinfo {volume} {8}},\
  \bibinfo {pages} {1901003} (\bibinfo {year} {2020})}\BibitemShut {NoStop}%
\bibitem [{\citenamefont {Frisk~Kockum}\ \emph {et~al.}(2019)\citenamefont
  {Frisk~Kockum}, \citenamefont {Miranowicz}, \citenamefont {De~Liberato},
  \citenamefont {Savasta},\ and\ \citenamefont
  {Nori}}]{frisk-kockum-ultrastrong-2019}%
  \BibitemOpen
  \bibfield  {author} {\bibinfo {author} {\bibfnamefont {A.}~\bibnamefont
  {Frisk~Kockum}}, \bibinfo {author} {\bibfnamefont {A.}~\bibnamefont
  {Miranowicz}}, \bibinfo {author} {\bibfnamefont {S.}~\bibnamefont
  {De~Liberato}}, \bibinfo {author} {\bibfnamefont {S.}~\bibnamefont
  {Savasta}},\ and\ \bibinfo {author} {\bibfnamefont {F.}~\bibnamefont
  {Nori}},\ }\bibfield  {title} {\bibinfo {title} {Ultrastrong coupling between
  light and matter},\ }\href {https://doi.org/10.1038/s42254-018-0006-2}
  {\bibfield  {journal} {\bibinfo  {journal} {Nature Reviews Physics}\ }\textbf
  {\bibinfo {volume} {1}},\ \bibinfo {pages} {19} (\bibinfo {year}
  {2019})}\BibitemShut {NoStop}%
\bibitem [{\citenamefont {Pelton}\ \emph {et~al.}(2019)\citenamefont {Pelton},
  \citenamefont {Storm},\ and\ \citenamefont {Leng}}]{pelton-strong-2019}%
  \BibitemOpen
  \bibfield  {author} {\bibinfo {author} {\bibfnamefont {M.}~\bibnamefont
  {Pelton}}, \bibinfo {author} {\bibfnamefont {S.~D.}\ \bibnamefont {Storm}},\
  and\ \bibinfo {author} {\bibfnamefont {H.}~\bibnamefont {Leng}},\ }\bibfield
  {title} {\bibinfo {title} {Strong coupling of emitters to single plasmonic
  nanoparticles: exciton-induced transparency and rabi splitting},\ }\href
  {https://doi.org/10.1039/C9NR05044B} {\bibfield  {journal} {\bibinfo
  {journal} {Nanoscale}\ }\textbf {\bibinfo {volume} {11}},\ \bibinfo {pages}
  {14540} (\bibinfo {year} {2019})}\BibitemShut {NoStop}%
\bibitem [{\citenamefont {Forn-D\'{\i}az}\ \emph {et~al.}(2019)\citenamefont
  {Forn-D\'{\i}az}, \citenamefont {Lamata}, \citenamefont {Rico}, \citenamefont
  {Kono},\ and\ \citenamefont {Solano}}]{forn-ultrastrong-2019}%
  \BibitemOpen
  \bibfield  {author} {\bibinfo {author} {\bibfnamefont {P.}~\bibnamefont
  {Forn-D\'{\i}az}}, \bibinfo {author} {\bibfnamefont {L.}~\bibnamefont
  {Lamata}}, \bibinfo {author} {\bibfnamefont {E.}~\bibnamefont {Rico}},
  \bibinfo {author} {\bibfnamefont {J.}~\bibnamefont {Kono}},\ and\ \bibinfo
  {author} {\bibfnamefont {E.}~\bibnamefont {Solano}},\ }\bibfield  {title}
  {\bibinfo {title} {Ultrastrong coupling regimes of light-matter
  interaction},\ }\href {https://doi.org/10.1103/RevModPhys.91.025005}
  {\bibfield  {journal} {\bibinfo  {journal} {Rev. Mod. Phys.}\ }\textbf
  {\bibinfo {volume} {91}},\ \bibinfo {pages} {025005} (\bibinfo {year}
  {2019})}\BibitemShut {NoStop}%
\bibitem [{\citenamefont {Nakayama}\ \emph {et~al.}(2013)\citenamefont
  {Nakayama}, \citenamefont {Kameda}, \citenamefont {Kawase},\ and\
  \citenamefont {Kim}}]{nakayama-control-2013}%
  \BibitemOpen
  \bibfield  {author} {\bibinfo {author} {\bibfnamefont {M.}~\bibnamefont
  {Nakayama}}, \bibinfo {author} {\bibfnamefont {M.}~\bibnamefont {Kameda}},
  \bibinfo {author} {\bibfnamefont {T.}~\bibnamefont {Kawase}},\ and\ \bibinfo
  {author} {\bibfnamefont {D.}~\bibnamefont {Kim}},\ }\bibfield  {title}
  {\bibinfo {title} {Control of rabi-splitting energies of exciton polaritons
  in cui microcavities},\ }\href
  {https://doi.org/https://doi.org/10.1140/epjb/e2012-30503-6} {\bibfield
  {journal} {\bibinfo  {journal} {European Physical Journal B}\ }\textbf
  {\bibinfo {volume} {86}},\ \bibinfo {pages} {32} (\bibinfo {year}
  {2013})}\BibitemShut {NoStop}%
\bibitem [{\citenamefont {Wang}\ \emph {et~al.}(2016)\citenamefont {Wang},
  \citenamefont {Li}, \citenamefont {Chervy}, \citenamefont {Shalabney},
  \citenamefont {Azzini}, \citenamefont {Orgiu}, \citenamefont {Hutchison},
  \citenamefont {Genet}, \citenamefont {Samorì},\ and\ \citenamefont
  {Ebbesen}}]{wang-coherent-2016}%
  \BibitemOpen
  \bibfield  {author} {\bibinfo {author} {\bibfnamefont {S.}~\bibnamefont
  {Wang}}, \bibinfo {author} {\bibfnamefont {S.}~\bibnamefont {Li}}, \bibinfo
  {author} {\bibfnamefont {T.}~\bibnamefont {Chervy}}, \bibinfo {author}
  {\bibfnamefont {A.}~\bibnamefont {Shalabney}}, \bibinfo {author}
  {\bibfnamefont {S.}~\bibnamefont {Azzini}}, \bibinfo {author} {\bibfnamefont
  {E.}~\bibnamefont {Orgiu}}, \bibinfo {author} {\bibfnamefont {J.~A.}\
  \bibnamefont {Hutchison}}, \bibinfo {author} {\bibfnamefont {C.}~\bibnamefont
  {Genet}}, \bibinfo {author} {\bibfnamefont {P.}~\bibnamefont {Samorì}},\
  and\ \bibinfo {author} {\bibfnamefont {T.~W.}\ \bibnamefont {Ebbesen}},\
  }\bibfield  {title} {\bibinfo {title} {Coherent coupling of
  \uppercase{WS}$_2$ monolayers with metallic photonic nanostructures at room
  temperature},\ }\href {https://doi.org/10.1021/acs.nanolett.6b01475}
  {\bibfield  {journal} {\bibinfo  {journal} {Nano Letters}\ }\textbf {\bibinfo
  {volume} {16}},\ \bibinfo {pages} {4368} (\bibinfo {year}
  {2016})}\BibitemShut {NoStop}%
\bibitem [{\citenamefont {Xie}\ \emph {et~al.}(2020)\citenamefont {Xie},
  \citenamefont {Li}, \citenamefont {Chen}, \citenamefont {Chang},
  \citenamefont {Zhang}, \citenamefont {Yi},\ and\ \citenamefont
  {Wang}}]{xie-enhanced-2020}%
  \BibitemOpen
  \bibfield  {author} {\bibinfo {author} {\bibfnamefont {P.}~\bibnamefont
  {Xie}}, \bibinfo {author} {\bibfnamefont {D.}~\bibnamefont {Li}}, \bibinfo
  {author} {\bibfnamefont {Y.}~\bibnamefont {Chen}}, \bibinfo {author}
  {\bibfnamefont {P.}~\bibnamefont {Chang}}, \bibinfo {author} {\bibfnamefont
  {H.}~\bibnamefont {Zhang}}, \bibinfo {author} {\bibfnamefont
  {J.}~\bibnamefont {Yi}},\ and\ \bibinfo {author} {\bibfnamefont
  {W.}~\bibnamefont {Wang}},\ }\bibfield  {title} {\bibinfo {title} {Enhanced
  coherent interaction between monolayer \uppercase{WS}e$_2$ and film-coupled
  nanocube open cavity with suppressed incoherent damping pathway},\ }\href
  {https://doi.org/10.1103/PhysRevB.102.115430} {\bibfield  {journal} {\bibinfo
   {journal} {Phys. Rev. B}\ }\textbf {\bibinfo {volume} {102}},\ \bibinfo
  {pages} {115430} (\bibinfo {year} {2020})}\BibitemShut {NoStop}%
\bibitem [{\citenamefont {Shapochkin}\ \emph {et~al.}(2020)\citenamefont
  {Shapochkin}, \citenamefont {Lozhkin}, \citenamefont {Solovev}, \citenamefont
  {Efimov}, \citenamefont {Eliseev}, \citenamefont {Lovtcius},\ and\
  \citenamefont {Kapitonov}}]{shapochkin-light-induced-2020}%
  \BibitemOpen
  \bibfield  {author} {\bibinfo {author} {\bibfnamefont {P.~Y.}\ \bibnamefont
  {Shapochkin}}, \bibinfo {author} {\bibfnamefont {M.~S.}\ \bibnamefont
  {Lozhkin}}, \bibinfo {author} {\bibfnamefont {I.~A.}\ \bibnamefont
  {Solovev}}, \bibinfo {author} {\bibfnamefont {Y.~P.}\ \bibnamefont {Efimov}},
  \bibinfo {author} {\bibfnamefont {S.~A.}\ \bibnamefont {Eliseev}}, \bibinfo
  {author} {\bibfnamefont {V.~A.}\ \bibnamefont {Lovtcius}},\ and\ \bibinfo
  {author} {\bibfnamefont {Y.~V.}\ \bibnamefont {Kapitonov}},\ }\bibfield
  {title} {\bibinfo {title} {Light-induced transition between the strong and
  weak coupling regimes in planar waveguide with
  \uppercase{G}a\uppercase{A}s/\uppercase{A}l\uppercase{G}a\uppercase{A}s
  quantum well},\ }\href {https://doi.org/10.1063/1.5141362} {\bibfield
  {journal} {\bibinfo  {journal} {Applied Physics Letters}\ }\textbf {\bibinfo
  {volume} {116}},\ \bibinfo {pages} {081102} (\bibinfo {year}
  {2020})}\BibitemShut {NoStop}%
\bibitem [{\citenamefont {G\'{o}mez}\ \emph {et~al.}(2021)\citenamefont
  {G\'{o}mez}, \citenamefont {Shi}, \citenamefont {Oshikiri}, \citenamefont
  {Roberts},\ and\ \citenamefont {Misawa}}]{gomez-near-perfect-2021}%
  \BibitemOpen
  \bibfield  {author} {\bibinfo {author} {\bibfnamefont {D.~E.}\ \bibnamefont
  {G\'{o}mez}}, \bibinfo {author} {\bibfnamefont {X.}~\bibnamefont {Shi}},
  \bibinfo {author} {\bibfnamefont {T.}~\bibnamefont {Oshikiri}}, \bibinfo
  {author} {\bibfnamefont {A.}~\bibnamefont {Roberts}},\ and\ \bibinfo {author}
  {\bibfnamefont {H.}~\bibnamefont {Misawa}},\ }\bibfield  {title} {\bibinfo
  {title} {Near-perfect absorption of light by coherent plasmon–exciton
  states},\ }\href {https://doi.org/10.1021/acs.nanolett.1c00389} {\bibfield
  {journal} {\bibinfo  {journal} {Nano Letters}\ }\textbf {\bibinfo {volume}
  {21}},\ \bibinfo {pages} {3864} (\bibinfo {year} {2021})},\ \bibinfo {note}
  {pMID: 33939440}\BibitemShut {NoStop}%
\bibitem [{\citenamefont {Leung}\ \emph {et~al.}(1994)\citenamefont {Leung},
  \citenamefont {Liu},\ and\ \citenamefont {Young}}]{leung-completeness-1994}%
  \BibitemOpen
  \bibfield  {author} {\bibinfo {author} {\bibfnamefont {P.~T.}\ \bibnamefont
  {Leung}}, \bibinfo {author} {\bibfnamefont {S.~Y.}\ \bibnamefont {Liu}},\
  and\ \bibinfo {author} {\bibfnamefont {K.}~\bibnamefont {Young}},\ }\bibfield
   {title} {{\selectlanguage {en}\bibinfo {title} {Completeness and
  orthogonality of quasinormal modes in leaky optical cavities}},\ }\href
  {https://doi.org/10.1103/PhysRevA.49.3057} {\bibfield  {journal} {\bibinfo
  {journal} {Physical Review A}\ }\textbf {\bibinfo {volume} {49}},\ \bibinfo
  {pages} {3057} (\bibinfo {year} {1994})}\BibitemShut {NoStop}%
\bibitem [{\citenamefont {Kristensen}\ \emph {et~al.}(2012)\citenamefont
  {Kristensen}, \citenamefont {Van~Vlack},\ and\ \citenamefont
  {Hughes}}]{kristensen-generalized-2012}%
  \BibitemOpen
  \bibfield  {author} {\bibinfo {author} {\bibfnamefont {P.~T.}\ \bibnamefont
  {Kristensen}}, \bibinfo {author} {\bibfnamefont {C.}~\bibnamefont
  {Van~Vlack}},\ and\ \bibinfo {author} {\bibfnamefont {S.}~\bibnamefont
  {Hughes}},\ }\bibfield  {title} {{\selectlanguage {en}\bibinfo {title}
  {Generalized effective mode volume for leaky optical cavities}},\ }\href
  {https://doi.org/10.1364/OL.37.001649} {\bibfield  {journal} {\bibinfo
  {journal} {Optics Letters}\ }\textbf {\bibinfo {volume} {37}},\ \bibinfo
  {pages} {1649} (\bibinfo {year} {2012})}\BibitemShut {NoStop}%
\bibitem [{\citenamefont {Lawless}\ \emph {et~al.}(0)\citenamefont {Lawless},
  \citenamefont {Hrelescu}, \citenamefont {Elliott}, \citenamefont {Peters},
  \citenamefont {McEvoy},\ and\ \citenamefont
  {Bradley}}]{lawless-influence-2020}%
  \BibitemOpen
  \bibfield  {author} {\bibinfo {author} {\bibfnamefont {J.}~\bibnamefont
  {Lawless}}, \bibinfo {author} {\bibfnamefont {C.}~\bibnamefont {Hrelescu}},
  \bibinfo {author} {\bibfnamefont {C.}~\bibnamefont {Elliott}}, \bibinfo
  {author} {\bibfnamefont {L.}~\bibnamefont {Peters}}, \bibinfo {author}
  {\bibfnamefont {N.}~\bibnamefont {McEvoy}},\ and\ \bibinfo {author}
  {\bibfnamefont {A.~L.}\ \bibnamefont {Bradley}},\ }\bibfield  {title}
  {\bibinfo {title} {Influence of gold nano-bipyramid dimensions on strong
  coupling with excitons of monolayer \uppercase{M}o\uppercase{S}$_2$},\ }\href
  {https://doi.org/10.1021/acsami.0c09261} {\bibfield  {journal} {\bibinfo
  {journal} {ACS Applied Materials \& Interfaces}\ }\textbf {\bibinfo {volume}
  {0}},\ \bibinfo {pages} {null} (\bibinfo {year} {0})}\BibitemShut {NoStop}%
\bibitem [{\citenamefont {Tserkezis}\ \emph {et~al.}(2020)\citenamefont
  {Tserkezis}, \citenamefont {Fern{\'{a}}ndez-Dom{\'{\i}}nguez}, \citenamefont
  {Gon{\c{c}}alves}, \citenamefont {Todisco}, \citenamefont {Cox},
  \citenamefont {Busch}, \citenamefont {Stenger}, \citenamefont {Bozhevolnyi},
  \citenamefont {Mortensen},\ and\ \citenamefont
  {Wolff}}]{tserkezis-applicability-2020}%
  \BibitemOpen
  \bibfield  {author} {\bibinfo {author} {\bibfnamefont {C.}~\bibnamefont
  {Tserkezis}}, \bibinfo {author} {\bibfnamefont {A.~I.}\ \bibnamefont
  {Fern{\'{a}}ndez-Dom{\'{\i}}nguez}}, \bibinfo {author} {\bibfnamefont
  {P.~A.~D.}\ \bibnamefont {Gon{\c{c}}alves}}, \bibinfo {author} {\bibfnamefont
  {F.}~\bibnamefont {Todisco}}, \bibinfo {author} {\bibfnamefont {J.~D.}\
  \bibnamefont {Cox}}, \bibinfo {author} {\bibfnamefont {K.}~\bibnamefont
  {Busch}}, \bibinfo {author} {\bibfnamefont {N.}~\bibnamefont {Stenger}},
  \bibinfo {author} {\bibfnamefont {S.~I.}\ \bibnamefont {Bozhevolnyi}},
  \bibinfo {author} {\bibfnamefont {N.~A.}\ \bibnamefont {Mortensen}},\ and\
  \bibinfo {author} {\bibfnamefont {C.}~\bibnamefont {Wolff}},\ }\bibfield
  {title} {\bibinfo {title} {On the applicability of quantum-optical concepts
  in strong-coupling nanophotonics},\ }\href
  {https://doi.org/10.1088/1361-6633/aba348} {\bibfield  {journal} {\bibinfo
  {journal} {Reports on Progress in Physics}\ }\textbf {\bibinfo {volume}
  {83}},\ \bibinfo {pages} {082401} (\bibinfo {year} {2020})}\BibitemShut
  {NoStop}%
\bibitem [{\citenamefont {Franke}\ \emph {et~al.}(2019)\citenamefont {Franke},
  \citenamefont {Hughes}, \citenamefont {Dezfouli}, \citenamefont {Kristensen},
  \citenamefont {Busch}, \citenamefont {Knorr},\ and\ \citenamefont
  {Richter}}]{franke-quantization-2019}%
  \BibitemOpen
  \bibfield  {author} {\bibinfo {author} {\bibfnamefont {S.}~\bibnamefont
  {Franke}}, \bibinfo {author} {\bibfnamefont {S.}~\bibnamefont {Hughes}},
  \bibinfo {author} {\bibfnamefont {M.~K.}\ \bibnamefont {Dezfouli}}, \bibinfo
  {author} {\bibfnamefont {P.~T.}\ \bibnamefont {Kristensen}}, \bibinfo
  {author} {\bibfnamefont {K.}~\bibnamefont {Busch}}, \bibinfo {author}
  {\bibfnamefont {A.}~\bibnamefont {Knorr}},\ and\ \bibinfo {author}
  {\bibfnamefont {M.}~\bibnamefont {Richter}},\ }\bibfield  {title} {\bibinfo
  {title} {Quantization of quasinormal modes for open cavities and plasmonic
  cavity quantum electrodynamics},\ }\href
  {https://doi.org/10.1103/PhysRevLett.122.213901} {\bibfield  {journal}
  {\bibinfo  {journal} {Phys. Rev. Lett.}\ }\textbf {\bibinfo {volume} {122}},\
  \bibinfo {pages} {213901} (\bibinfo {year} {2019})}\BibitemShut {NoStop}%
\bibitem [{\citenamefont {Franke}\ \emph {et~al.}(2020)\citenamefont {Franke},
  \citenamefont {Richter}, \citenamefont {Ren}, \citenamefont {Knorr},\ and\
  \citenamefont {Hughes}}]{PhysRevResearch.2.033456}%
  \BibitemOpen
  \bibfield  {author} {\bibinfo {author} {\bibfnamefont {S.}~\bibnamefont
  {Franke}}, \bibinfo {author} {\bibfnamefont {M.}~\bibnamefont {Richter}},
  \bibinfo {author} {\bibfnamefont {J.}~\bibnamefont {Ren}}, \bibinfo {author}
  {\bibfnamefont {A.}~\bibnamefont {Knorr}},\ and\ \bibinfo {author}
  {\bibfnamefont {S.}~\bibnamefont {Hughes}},\ }\bibfield  {title} {\bibinfo
  {title} {Quantized quasinormal-mode description of nonlinear cavity-{QED}
  effects from coupled resonators with a fano-like resonance},\ }\href
  {https://doi.org/10.1103/PhysRevResearch.2.033456} {\bibfield  {journal}
  {\bibinfo  {journal} {Phys. Rev. Research}\ }\textbf {\bibinfo {volume}
  {2}},\ \bibinfo {pages} {033456} (\bibinfo {year} {2020})}\BibitemShut
  {NoStop}%
\bibitem [{\citenamefont {Zhu}\ \emph {et~al.}(2016{\natexlab{b}})\citenamefont
  {Zhu}, \citenamefont {Esteban}, \citenamefont {Borisov}, \citenamefont
  {Baumberg}, \citenamefont {Nordlander}, \citenamefont {Lezec}, \citenamefont
  {Aizpurua},\ and\ \citenamefont {Crozier}}]{Zhu2016}%
  \BibitemOpen
  \bibfield  {author} {\bibinfo {author} {\bibfnamefont {W.}~\bibnamefont
  {Zhu}}, \bibinfo {author} {\bibfnamefont {R.}~\bibnamefont {Esteban}},
  \bibinfo {author} {\bibfnamefont {A.~G.}\ \bibnamefont {Borisov}}, \bibinfo
  {author} {\bibfnamefont {J.~J.}\ \bibnamefont {Baumberg}}, \bibinfo {author}
  {\bibfnamefont {P.}~\bibnamefont {Nordlander}}, \bibinfo {author}
  {\bibfnamefont {H.~J.}\ \bibnamefont {Lezec}}, \bibinfo {author}
  {\bibfnamefont {J.}~\bibnamefont {Aizpurua}},\ and\ \bibinfo {author}
  {\bibfnamefont {K.~B.}\ \bibnamefont {Crozier}},\ }\bibfield  {title}
  {\bibinfo {title} {Quantum mechanical effects in plasmonic structures with
  subnanometre gaps},\ }\bibfield  {journal} {\bibinfo  {journal} {Nature
  Communications}\ }\textbf {\bibinfo {volume} {7}},\ \href
  {https://doi.org/10.1038/ncomms11495} {10.1038/ncomms11495} (\bibinfo {year}
  {2016}{\natexlab{b}})\BibitemShut {NoStop}%
\bibitem [{\citenamefont {Dezfouli}\ \emph {et~al.}(2017)\citenamefont
  {Dezfouli}, \citenamefont {Tserkezis}, \citenamefont {Mortensen},\ and\
  \citenamefont {Hughes}}]{KamandarDezfouli2017}%
  \BibitemOpen
  \bibfield  {author} {\bibinfo {author} {\bibfnamefont {M.~K.}\ \bibnamefont
  {Dezfouli}}, \bibinfo {author} {\bibfnamefont {C.}~\bibnamefont {Tserkezis}},
  \bibinfo {author} {\bibfnamefont {N.~A.}\ \bibnamefont {Mortensen}},\ and\
  \bibinfo {author} {\bibfnamefont {S.}~\bibnamefont {Hughes}},\ }\bibfield
  {title} {\bibinfo {title} {Nonlocal quasinormal modes for arbitrarily shaped
  three-dimensional plasmonic resonators},\ }\href
  {https://doi.org/10.1364/optica.4.001503} {\bibfield  {journal} {\bibinfo
  {journal} {Optica}\ }\textbf {\bibinfo {volume} {4}},\ \bibinfo {pages}
  {1503} (\bibinfo {year} {2017})}\BibitemShut {NoStop}%
\bibitem [{\citenamefont {Haug}\ and\ \citenamefont
  {Koch}(2009)}]{haug2009quantum}%
  \BibitemOpen
  \bibfield  {author} {\bibinfo {author} {\bibfnamefont {H.}~\bibnamefont
  {Haug}}\ and\ \bibinfo {author} {\bibfnamefont {S.~W.}\ \bibnamefont
  {Koch}},\ } {} {\emph {\bibinfo {title} {Quantum Theory of the
  Optical and Electronic Properties of Semiconductors}}}\ (\bibinfo
  {publisher} {World Scientific},\ \bibinfo {year} {2009})\BibitemShut
  {NoStop}%
\bibitem [{\citenamefont {Kira}\ and\ \citenamefont
  {Koch}(2006)}]{kira2006many}%
  \BibitemOpen
  \bibfield  {author} {\bibinfo {author} {\bibfnamefont {M.}~\bibnamefont
  {Kira}}\ and\ \bibinfo {author} {\bibfnamefont {S.~W.}\ \bibnamefont
  {Koch}},\ }\bibfield  {title} {\bibinfo {title} {Many-body correlations and
  excitonic effects in semiconductor spectroscopy},\ }{} {\bibfield
  {journal} {\bibinfo  {journal} {Progress in quantum electronics}\ }\textbf
  {\bibinfo {volume} {30}},\ \bibinfo {pages} {155} (\bibinfo {year}
  {2006})}\BibitemShut {NoStop}%
\bibitem [{\citenamefont {Trolle}\ \emph {et~al.}(2017)\citenamefont {Trolle},
  \citenamefont {Pedersen},\ and\ \citenamefont
  {V{\'e}niard}}]{trolle2017model}%
  \BibitemOpen
  \bibfield  {author} {\bibinfo {author} {\bibfnamefont {M.~L.}\ \bibnamefont
  {Trolle}}, \bibinfo {author} {\bibfnamefont {T.~G.}\ \bibnamefont
  {Pedersen}},\ and\ \bibinfo {author} {\bibfnamefont {V.}~\bibnamefont
  {V{\'e}niard}},\ }\bibfield  {title} {\bibinfo {title} {Model dielectric
  function for 2d semiconductors including substrate screening},\ }
  {} {\bibfield  {journal} {\bibinfo  {journal} {Scientific reports}\ }\textbf
  {\bibinfo {volume} {7}},\ \bibinfo {pages} {39844} (\bibinfo {year}
  {2017})}\BibitemShut {NoStop}%
\bibitem [{\citenamefont {Martin}\ \emph {et~al.}(2020)\citenamefont {Martin},
  \citenamefont {Horng}, \citenamefont {Ruth}, \citenamefont {Paik},
  \citenamefont {Wentzel}, \citenamefont {Deng},\ and\ \citenamefont
  {Cundiff}}]{martin-encapsulation-2020}%
  \BibitemOpen
  \bibfield  {author} {\bibinfo {author} {\bibfnamefont {E.~W.}\ \bibnamefont
  {Martin}}, \bibinfo {author} {\bibfnamefont {J.}~\bibnamefont {Horng}},
  \bibinfo {author} {\bibfnamefont {H.~G.}\ \bibnamefont {Ruth}}, \bibinfo
  {author} {\bibfnamefont {E.}~\bibnamefont {Paik}}, \bibinfo {author}
  {\bibfnamefont {M.-H.}\ \bibnamefont {Wentzel}}, \bibinfo {author}
  {\bibfnamefont {H.}~\bibnamefont {Deng}},\ and\ \bibinfo {author}
  {\bibfnamefont {S.~T.}\ \bibnamefont {Cundiff}},\ }\bibfield  {title}
  {\bibinfo {title} {Encapsulation narrows and preserves the excitonic
  homogeneous linewidth of exfoliated monolayer
  \uppercase{M}o\uppercase{S}e$_2$},\ }\href
  {https://doi.org/10.1103/PhysRevApplied.14.021002} {\bibfield  {journal}
  {\bibinfo  {journal} {Phys. Rev. Applied}\ }\textbf {\bibinfo {volume}
  {14}},\ \bibinfo {pages} {021002} (\bibinfo {year} {2020})}\BibitemShut
  {NoStop}%
\bibitem [{\citenamefont {Christiansen}\ \emph {et~al.}(2017)\citenamefont
  {Christiansen}, \citenamefont {Selig}, \citenamefont {Bergh\"auser},
  \citenamefont {Schmidt}, \citenamefont {Niehues}, \citenamefont {Schneider},
  \citenamefont {Arora}, \citenamefont {de~Vasconcellos}, \citenamefont
  {Bratschitsch}, \citenamefont {Mali\'{c}},\ and\ \citenamefont
  {Knorr}}]{christiansen2017phonon}%
  \BibitemOpen
  \bibfield  {author} {\bibinfo {author} {\bibfnamefont {D.}~\bibnamefont
  {Christiansen}}, \bibinfo {author} {\bibfnamefont {M.}~\bibnamefont {Selig}},
  \bibinfo {author} {\bibfnamefont {G.}~\bibnamefont {Bergh\"auser}}, \bibinfo
  {author} {\bibfnamefont {R.}~\bibnamefont {Schmidt}}, \bibinfo {author}
  {\bibfnamefont {I.}~\bibnamefont {Niehues}}, \bibinfo {author} {\bibfnamefont
  {R.}~\bibnamefont {Schneider}}, \bibinfo {author} {\bibfnamefont
  {A.}~\bibnamefont {Arora}}, \bibinfo {author} {\bibfnamefont {S.~M.}\
  \bibnamefont {de~Vasconcellos}}, \bibinfo {author} {\bibfnamefont
  {R.}~\bibnamefont {Bratschitsch}}, \bibinfo {author} {\bibfnamefont
  {E.}~\bibnamefont {Mali\'{c}}},\ and\ \bibinfo {author} {\bibfnamefont
  {A.}~\bibnamefont {Knorr}},\ }\bibfield  {title} {\bibinfo {title} {Phonon
  sidebands in monolayer transition metal dichalcogenides},\ } {}
  {\bibfield  {journal} {\bibinfo  {journal} {Physical Review Letters}\
  }\textbf {\bibinfo {volume} {119}},\ \bibinfo {pages} {187402} (\bibinfo
  {year} {2017})}\BibitemShut {NoStop}%
\bibitem [{\citenamefont {Brem}\ \emph {et~al.}(2019)\citenamefont {Brem},
  \citenamefont {Zipfel}, \citenamefont {Selig}, \citenamefont {Raja},
  \citenamefont {Waldecker}, \citenamefont {Ziegler}, \citenamefont
  {Taniguchi}, \citenamefont {Watanabe}, \citenamefont {Chernikov},\ and\
  \citenamefont {Mali\'{c}}}]{brem2019intrinsic}%
  \BibitemOpen
  \bibfield  {author} {\bibinfo {author} {\bibfnamefont {S.}~\bibnamefont
  {Brem}}, \bibinfo {author} {\bibfnamefont {J.}~\bibnamefont {Zipfel}},
  \bibinfo {author} {\bibfnamefont {M.}~\bibnamefont {Selig}}, \bibinfo
  {author} {\bibfnamefont {A.}~\bibnamefont {Raja}}, \bibinfo {author}
  {\bibfnamefont {L.}~\bibnamefont {Waldecker}}, \bibinfo {author}
  {\bibfnamefont {J.~D.}\ \bibnamefont {Ziegler}}, \bibinfo {author}
  {\bibfnamefont {T.}~\bibnamefont {Taniguchi}}, \bibinfo {author}
  {\bibfnamefont {K.}~\bibnamefont {Watanabe}}, \bibinfo {author}
  {\bibfnamefont {A.}~\bibnamefont {Chernikov}},\ and\ \bibinfo {author}
  {\bibfnamefont {E.}~\bibnamefont {Mali\'{c}}},\ }\bibfield  {title} {\bibinfo
  {title} {Intrinsic lifetime of higher excitonic states in tungsten diselenide
  monolayers},\ } {} {\bibfield  {journal} {\bibinfo  {journal}
  {Nanoscale}\ } (\bibinfo {year} {2019})}\BibitemShut {NoStop}%
\bibitem [{\citenamefont {Geick}\ \emph {et~al.}(1966)\citenamefont {Geick},
  \citenamefont {Perry},\ and\ \citenamefont {Rupprecht}}]{geick1966normal}%
  \BibitemOpen
  \bibfield  {author} {\bibinfo {author} {\bibfnamefont {R.}~\bibnamefont
  {Geick}}, \bibinfo {author} {\bibfnamefont {C.}~\bibnamefont {Perry}},\ and\
  \bibinfo {author} {\bibfnamefont {G.}~\bibnamefont {Rupprecht}},\ }\bibfield
  {title} {\bibinfo {title} {Normal modes in hexagonal boron nitride},\
  } {} {\bibfield  {journal} {\bibinfo  {journal} {Physical Review}\
  }\textbf {\bibinfo {volume} {146}},\ \bibinfo {pages} {543} (\bibinfo {year}
  {1966})}\BibitemShut {NoStop}%
\bibitem [{\citenamefont {Palik}(1998)}]{palikgold}%
  \BibitemOpen
  \bibfield  {author} {\bibinfo {author} {\bibfnamefont {E.}~\bibnamefont
  {Palik}},\ } {} {\emph {\bibinfo {title} {Handbook of Optical
  Constants of Solids I - III}}}\ (\bibinfo  {publisher} {Academic Press},\
  \bibinfo {year} {1998})\BibitemShut {NoStop}%
\bibitem [{com()}]{comsol}%
  \BibitemOpen
  \href {www.comsol.com} {\bibinfo {title} {Comsol multiphysics v. 5.3}},\
  \bibinfo {note} {\uppercase{COMSOL} Inc.}\BibitemShut {Stop}%
\bibitem [{\citenamefont {Bai}\ \emph {et~al.}(2013)\citenamefont {Bai},
  \citenamefont {Perrin}, \citenamefont {Sauvan}, \citenamefont {Hugonin},\
  and\ \citenamefont {Lalanne}}]{bai-efficient-2013}%
  \BibitemOpen
  \bibfield  {author} {\bibinfo {author} {\bibfnamefont {Q.}~\bibnamefont
  {Bai}}, \bibinfo {author} {\bibfnamefont {M.}~\bibnamefont {Perrin}},
  \bibinfo {author} {\bibfnamefont {C.}~\bibnamefont {Sauvan}}, \bibinfo
  {author} {\bibfnamefont {J.-P.}\ \bibnamefont {Hugonin}},\ and\ \bibinfo
  {author} {\bibfnamefont {P.}~\bibnamefont {Lalanne}},\ }\bibfield  {title}
  {{\selectlanguage {EN}\bibinfo {title} {Efficient and intuitive method for
  the analysis of light scattering by a resonant nanostructure}},\ }\href
  {https://doi.org/10.1364/OE.21.027371} {\bibfield  {journal} {\bibinfo
  {journal} {Optics Express}\ }\textbf {\bibinfo {volume} {21}},\ \bibinfo
  {pages} {27371} (\bibinfo {year} {2013})}\BibitemShut {NoStop}%
\bibitem [{\citenamefont {Carlson}\ and\ \citenamefont
  {Hughes}(2020)}]{carlson-dissipative-2020}%
  \BibitemOpen
  \bibfield  {author} {\bibinfo {author} {\bibfnamefont {C.}~\bibnamefont
  {Carlson}}\ and\ \bibinfo {author} {\bibfnamefont {S.}~\bibnamefont
  {Hughes}},\ }\bibfield  {title} {\bibinfo {title} {Dissipative modes, purcell
  factors, and directional beta factors in gold bowtie nanoantenna
  structures},\ }\href {https://doi.org/10.1103/PhysRevB.102.155301} {\bibfield
   {journal} {\bibinfo  {journal} {Phys. Rev. B}\ }\textbf {\bibinfo {volume}
  {102}},\ \bibinfo {pages} {155301} (\bibinfo {year} {2020})}\BibitemShut
  {NoStop}%
\bibitem [{\citenamefont {de~Lasson}\ \emph {et~al.}(2015)\citenamefont
  {de~Lasson}, \citenamefont {Kristensen}, \citenamefont {M{\o}rk},\ and\
  \citenamefont {Gregersen}}]{RosenkrantzdeLasson2015}%
  \BibitemOpen
  \bibfield  {author} {\bibinfo {author} {\bibfnamefont {J.~R.}\ \bibnamefont
  {de~Lasson}}, \bibinfo {author} {\bibfnamefont {P.~T.}\ \bibnamefont
  {Kristensen}}, \bibinfo {author} {\bibfnamefont {J.}~\bibnamefont
  {M{\o}rk}},\ and\ \bibinfo {author} {\bibfnamefont {N.}~\bibnamefont
  {Gregersen}},\ }\bibfield  {title} {\bibinfo {title} {Semianalytical
  quasi-normal mode theory for the local density of states in coupled photonic
  crystal cavity{\textendash}waveguide structures},\ }\href
  {https://doi.org/10.1364/ol.40.005790} {\bibfield  {journal} {\bibinfo
  {journal} {Optics Letters}\ }\textbf {\bibinfo {volume} {40}},\ \bibinfo
  {pages} {5790} (\bibinfo {year} {2015})}\BibitemShut {NoStop}%
\bibitem [{\citenamefont {Kamandar~Dezfouli}\ \emph {et~al.}(2017)\citenamefont
  {Kamandar~Dezfouli}, \citenamefont {Gordon},\ and\ \citenamefont
  {Hughes}}]{2017PRA-hybrid}%
  \BibitemOpen
  \bibfield  {author} {\bibinfo {author} {\bibfnamefont {M.}~\bibnamefont
  {Kamandar~Dezfouli}}, \bibinfo {author} {\bibfnamefont {R.}~\bibnamefont
  {Gordon}},\ and\ \bibinfo {author} {\bibfnamefont {S.}~\bibnamefont
  {Hughes}},\ }\bibfield  {title} {\bibinfo {title} {Modal theory of modified
  spontaneous emission of a quantum emitter in a hybrid plasmonic
  photonic-crystal cavity system},\ }\href
  {https://doi.org/10.1103/PhysRevA.95.013846} {\bibfield  {journal} {\bibinfo
  {journal} {Phys. Rev. A}\ }\textbf {\bibinfo {volume} {95}},\ \bibinfo
  {pages} {013846} (\bibinfo {year} {2017})}\BibitemShut {NoStop}%
\bibitem [{\citenamefont {Ge}\ \emph {et~al.}(2014{\natexlab{a}})\citenamefont
  {Ge}, \citenamefont {Kristensen}, \citenamefont {Young},\ and\ \citenamefont
  {Hughes}}]{ge-quasinormal-2014}%
  \BibitemOpen
  \bibfield  {author} {\bibinfo {author} {\bibfnamefont {R.-C.}\ \bibnamefont
  {Ge}}, \bibinfo {author} {\bibfnamefont {P.~T.}\ \bibnamefont {Kristensen}},
  \bibinfo {author} {\bibfnamefont {J.~F.}\ \bibnamefont {Young}},\ and\
  \bibinfo {author} {\bibfnamefont {S.}~\bibnamefont {Hughes}},\ }\bibfield
  {title} {\bibinfo {title} {Quasinormal mode approach to modelling
  light-emission and propagation in nanoplasmonics},\ }\href
  {https://doi.org/10.1088/1367-2630/16/11/113048} {\bibfield  {journal}
  {\bibinfo  {journal} {New Journal of Physics}\ }\textbf {\bibinfo {volume}
  {16}},\ \bibinfo {pages} {113048} (\bibinfo {year}
  {2014}{\natexlab{a}})}\BibitemShut {NoStop}%
\bibitem [{\citenamefont {Ren}\ \emph {et~al.}(2020)\citenamefont {Ren},
  \citenamefont {Franke}, \citenamefont {Knorr}, \citenamefont {Richter},\ and\
  \citenamefont {Hughes}}]{PhysRevB.101.205402}%
  \BibitemOpen
  \bibfield  {author} {\bibinfo {author} {\bibfnamefont {J.}~\bibnamefont
  {Ren}}, \bibinfo {author} {\bibfnamefont {S.}~\bibnamefont {Franke}},
  \bibinfo {author} {\bibfnamefont {A.}~\bibnamefont {Knorr}}, \bibinfo
  {author} {\bibfnamefont {M.}~\bibnamefont {Richter}},\ and\ \bibinfo {author}
  {\bibfnamefont {S.}~\bibnamefont {Hughes}},\ }\bibfield  {title} {\bibinfo
  {title} {Near-field to far-field transformations of optical quasinormal modes
  and efficient calculation of quantized quasinormal modes for open cavities
  and plasmonic resonators},\ }\href
  {https://doi.org/10.1103/PhysRevB.101.205402} {\bibfield  {journal} {\bibinfo
   {journal} {Phys. Rev. B}\ }\textbf {\bibinfo {volume} {101}},\ \bibinfo
  {pages} {205402} (\bibinfo {year} {2020})}\BibitemShut {NoStop}%
\bibitem [{\citenamefont {Binkowski}\ \emph {et~al.}(2020)\citenamefont
  {Binkowski}, \citenamefont {Betz}, \citenamefont {Colom}, \citenamefont
  {Hammerschmidt}, \citenamefont {Zschiedrich},\ and\ \citenamefont
  {Burger}}]{PhysRevB.102.035432}%
  \BibitemOpen
  \bibfield  {author} {\bibinfo {author} {\bibfnamefont {F.}~\bibnamefont
  {Binkowski}}, \bibinfo {author} {\bibfnamefont {F.}~\bibnamefont {Betz}},
  \bibinfo {author} {\bibfnamefont {R.}~\bibnamefont {Colom}}, \bibinfo
  {author} {\bibfnamefont {M.}~\bibnamefont {Hammerschmidt}}, \bibinfo {author}
  {\bibfnamefont {L.}~\bibnamefont {Zschiedrich}},\ and\ \bibinfo {author}
  {\bibfnamefont {S.}~\bibnamefont {Burger}},\ }\bibfield  {title} {\bibinfo
  {title} {Quasinormal mode expansion of optical far-field quantities},\ }\href
  {https://doi.org/10.1103/PhysRevB.102.035432} {\bibfield  {journal} {\bibinfo
   {journal} {Phys. Rev. B}\ }\textbf {\bibinfo {volume} {102}},\ \bibinfo
  {pages} {035432} (\bibinfo {year} {2020})}\BibitemShut {NoStop}%
\bibitem [{\citenamefont {Ge}\ \emph {et~al.}(2014{\natexlab{b}})\citenamefont
  {Ge}, \citenamefont {Kristensen}, \citenamefont {Young},\ and\ \citenamefont
  {Hughes}}]{GeNJP2014}%
  \BibitemOpen
  \bibfield  {author} {\bibinfo {author} {\bibfnamefont {R.-C.}\ \bibnamefont
  {Ge}}, \bibinfo {author} {\bibfnamefont {P.~T.}\ \bibnamefont {Kristensen}},
  \bibinfo {author} {\bibfnamefont {J.~F.}\ \bibnamefont {Young}},\ and\
  \bibinfo {author} {\bibfnamefont {S.}~\bibnamefont {Hughes}},\ }\bibfield
  {title} {\bibinfo {title} {Quasinormal mode approach to modelling
  light-emission and propagation in nanoplasmonics},\ } {} {\bibfield
   {journal} {\bibinfo  {journal} {New J. Phys.}\ }\textbf {\bibinfo {volume}
  {16}},\ \bibinfo {pages} {113048} (\bibinfo {year}
  {2014}{\natexlab{b}})}\BibitemShut {NoStop}%
\bibitem [{\citenamefont {Şendur}\ and\ \citenamefont
  {Baran}(2009)}]{sendur-near-2009}%
  \BibitemOpen
  \bibfield  {author} {\bibinfo {author} {\bibfnamefont {K.}~\bibnamefont
  {Şendur}}\ and\ \bibinfo {author} {\bibfnamefont {E.}~\bibnamefont
  {Baran}},\ }\bibfield  {title} {\bibinfo {title} {Near-field optical power
  transmission of dipole nano-antennas},\ }\href
  {https://doi.org/10.1007/s00340-009-3505-0} {\bibfield  {journal} {\bibinfo
  {journal} {Appl. Phys. B}\ }\textbf {\bibinfo {volume} {96}},\ \bibinfo
  {pages} {325} (\bibinfo {year} {2009})}\BibitemShut {NoStop}%
\bibitem [{\citenamefont {Noda}\ \emph {et~al.}(2007)\citenamefont {Noda},
  \citenamefont {Fujita},\ and\ \citenamefont {Asano}}]{noda-spontaneous-2007}%
  \BibitemOpen
  \bibfield  {author} {\bibinfo {author} {\bibfnamefont {S.}~\bibnamefont
  {Noda}}, \bibinfo {author} {\bibfnamefont {M.}~\bibnamefont {Fujita}},\ and\
  \bibinfo {author} {\bibfnamefont {T.}~\bibnamefont {Asano}},\ }\bibfield
  {title} {\bibinfo {title} {Spontaneous-emission control by photonic crystals
  and nanocavities},\ }\href {https://doi.org/10.1038/nphoton.2007.141}
  {\bibfield  {journal} {\bibinfo  {journal} {Nature Photonics}\ }\textbf
  {\bibinfo {volume} {1}},\ \bibinfo {pages} {449–458} (\bibinfo {year}
  {2007})}\BibitemShut {NoStop}%
\bibitem [{\citenamefont {Checoury}\ \emph {et~al.}(2010)\citenamefont
  {Checoury}, \citenamefont {Han}, \citenamefont {El~Kurdi},\ and\
  \citenamefont {Boucaud}}]{checoury-deterministic-2010}%
  \BibitemOpen
  \bibfield  {author} {\bibinfo {author} {\bibfnamefont {X.}~\bibnamefont
  {Checoury}}, \bibinfo {author} {\bibfnamefont {Z.}~\bibnamefont {Han}},
  \bibinfo {author} {\bibfnamefont {M.}~\bibnamefont {El~Kurdi}},\ and\
  \bibinfo {author} {\bibfnamefont {P.}~\bibnamefont {Boucaud}},\ }\bibfield
  {title} {\bibinfo {title} {Deterministic measurement of the purcell factor in
  microcavities through raman emission},\ }\href
  {https://doi.org/10.1103/PhysRevA.81.033832} {\bibfield  {journal} {\bibinfo
  {journal} {Phys. Rev. A}\ }\textbf {\bibinfo {volume} {81}},\ \bibinfo
  {pages} {033832} (\bibinfo {year} {2010})}\BibitemShut {NoStop}%
\bibitem [{\citenamefont {Ge}\ and\ \citenamefont
  {Hughes}(2015)}]{PhysRevB.92.205420}%
  \BibitemOpen
  \bibfield  {author} {\bibinfo {author} {\bibfnamefont {R.-C.}\ \bibnamefont
  {Ge}}\ and\ \bibinfo {author} {\bibfnamefont {S.}~\bibnamefont {Hughes}},\
  }\bibfield  {title} {\bibinfo {title} {Quantum dynamics of two quantum dots
  coupled through localized plasmons: An intuitive and accurate quantum optics
  approach using quasinormal modes},\ }\href
  {https://doi.org/10.1103/PhysRevB.92.205420} {\bibfield  {journal} {\bibinfo
  {journal} {Phys. Rev. B}\ }\textbf {\bibinfo {volume} {92}},\ \bibinfo
  {pages} {205420} (\bibinfo {year} {2015})}\BibitemShut {NoStop}%
\bibitem [{\citenamefont {Feist}\ \emph {et~al.}(2020)\citenamefont {Feist},
  \citenamefont {Fern{\'{a}}ndez-Dom{\'{\i}}nguez},\ and\ \citenamefont
  {Garc{\'{\i}}a-Vidal}}]{Feist2020}%
  \BibitemOpen
  \bibfield  {author} {\bibinfo {author} {\bibfnamefont {J.}~\bibnamefont
  {Feist}}, \bibinfo {author} {\bibfnamefont {A.~I.}\ \bibnamefont
  {Fern{\'{a}}ndez-Dom{\'{\i}}nguez}},\ and\ \bibinfo {author} {\bibfnamefont
  {F.~J.}\ \bibnamefont {Garc{\'{\i}}a-Vidal}},\ }\bibfield  {title} {\bibinfo
  {title} {Macroscopic {QED} for quantum nanophotonics: emitter-centered modes
  as a minimal basis for multiemitter problems},\ }\href
  {https://doi.org/10.1515/nanoph-2020-0451} {\bibfield  {journal} {\bibinfo
  {journal} {Nanophotonics}\ }\textbf {\bibinfo {volume} {10}},\ \bibinfo
  {pages} {477} (\bibinfo {year} {2020})}\BibitemShut {NoStop}%
\bibitem [{\citenamefont {Denning}\ \emph {et~al.}(2021)\citenamefont
  {Denning}, \citenamefont {Wubs}, \citenamefont {Stenger}, \citenamefont
  {Mork},\ and\ \citenamefont {Kristensen}}]{denning-quantum-2021}%
  \BibitemOpen
  \bibfield  {author} {\bibinfo {author} {\bibfnamefont {E.~V.}\ \bibnamefont
  {Denning}}, \bibinfo {author} {\bibfnamefont {M.}~\bibnamefont {Wubs}},
  \bibinfo {author} {\bibfnamefont {N.}~\bibnamefont {Stenger}}, \bibinfo
  {author} {\bibfnamefont {J.}~\bibnamefont {Mork}},\ and\ \bibinfo {author}
  {\bibfnamefont {P.~T.}\ \bibnamefont {Kristensen}},\ } {} {\bibinfo
  {title} {Quantum theory of two-dimensional materials coupled to
  electromagnetic resonators}} (\bibinfo {year} {2021}),\ 
  \href{https://arxiv.org/abs/2103.14488} {arXiv:2103.14488 [quant-ph]} 
\BibitemShut
  {NoStop}%
\end{thebibliography}%

\end{document}